\documentclass{jfm}

\usepackage{graphicx}
\usepackage{natbib}

\usepackage{graphics,epsfig}
\usepackage{color}

\def\build#1_#2^#3{\mathrel{\mathop{\kern 0pt#1}\limits_{#2}^{#3}}}

 \newcommand{\vct}[1]{{\mbox {\boldmath $#1$}}}
 \newcommand{\be}{\begin{equation}}
 \newcommand{\ee}{\end{equation}}

\ifCUPmtlplainloaded \else
  \checkfont{eurm10}
  \iffontfound
    \IfFileExists{upmath.sty}
      {\typeout{^^JFound AMS Euler Roman fonts on the system,
                   using the 'upmath' package.^^J}%
       \usepackage{upmath}}
      {\typeout{^^JFound AMS Euler Roman fonts on the system, but you
                   dont seem to have the}%
       \typeout{'upmath' package installed. JFM.cls can take advantage
                 of these fonts,^^Jif you use 'upmath' package.^^J}%
      }
  \else
  \fi
\fi

\ifCUPmtlplainloaded \else
  \checkfont{msam10}
  \iffontfound
    \IfFileExists{amssymb.sty}
      {\typeout{^^JFound AMS Symbol fonts on the system, using the
                'amssymb' package.^^J}%
       \usepackage{amssymb}%

      }{}
  \fi
\fi

\ifCUPmtlplainloaded \else
  \IfFileExists{amsbsy.sty}
    {\typeout{^^JFound the 'amsbsy' package on the system, using it.^^J}%
     \usepackage{amsbsy}}
    {}
\fi

\newsavebox{\astrutbox}
\sbox{\astrutbox}{\rule[-5pt]{0pt}{20pt}}

\title[Orientation dynamics of ellipsoids in turbulence]{Orientation dynamics of small, triaxial-ellipsoidal particles in isotropic turbulence}

\author[ ]%
{Laurent Chevillard$^1$\thanks{laurent.chevillard@ens-lyon.fr}, and Charles Meneveau$^{2}$\thanks{meneveau@jhu.edu}}

\affiliation{$^1$Laboratoire de Physique de l'\'Ecole Normale Sup\'erieure
de Lyon, CNRS, Universit\'e de Lyon, 46 all\'ee d'Italie F-69007 Lyon,
France \\
$^2$Department of Mechanical Engineering and Center
for Environmental and Applied Fluid Mechanics, The Johns Hopkins
University, 3400 N. Charles Street, Baltimore, MD 21218,
USA}
\pubyear{2013}
\volume{..}
\pagerange{..}
\date{..}
\begin{document}

\bibliographystyle{jfm}

\maketitle

\begin{abstract}
The orientation dynamics of  small anisotropic tracer particles in turbulent flows is studied using direct numerical simulation (DNS) and results are compared with Lagrangian stochastic models.  Generalizing earlier analysis for axisymmetric ellipsoidal particles \citep{Parsaetal12}, we measure the orientation statistics and rotation rates of general, triaxial ellipsoidal tracer particles using Lagrangian tracking in DNS of isotropic turbulence. Triaxial ellipsoids that are very long in one direction, very thin in another, and of
intermediate size in the third direction exhibit reduced rotation rates that are similar to those of rods in the ellipsoid's longest direction, while exhibiting increased rotation rates that are similar to those of axisymmetric discs in the thinnest direction. DNS results differ significantly from the case when the particle orientations are assumed to be statistically independent from the velocity gradient tensor. They are also different from predictions of a Gaussian process for the velocity gradient tensor, which does not provide realistic preferred vorticity-strain-rate tensor alignments. DNS results are also compared with a stochastic model for the velocity gradient tensor based on the recent fluid deformation approximation (RFDA).  Unlike the Gaussian model, the stochastic model accurately predicts the reduction in rotation rate in the longest direction of triaxial ellipsoids since this direction aligns with the flow's vorticity, with its rotation perpendicular to the vorticity being reduced.  For disc-like particles, or in directions perpendicular to the longest direction in triaxial particles, the model predicts {noticeably} smaller rotation rates than those observed in DNS, a behavior that can be understood based on the probability of vorticity orientation with the most contracting strain-rate eigen-direction in the model.  
\end{abstract}

\section{Introduction}
The fate of anisotropic particles in fluid flows is of considerable interest in the context of various applications, such as micro-organism locomotion \citep{PedleyKessler92,KochSubramanian11,SaintillanShelley07}, 
industrial  manufacturing processes such as paper-making \citep{Lundelletal11},
and natural phenomena such as ice crystal formation in clouds  \citep{PinskyCain98}.
In many of these applications, the flow is highly turbulent and the rotational dynamics, alignment trends and correlations  of anisotropic particles (such as fibers, discs or more general shapes) with the flow field become of considerable interest. For small tracer particles whose size is smaller than the Kolmogorov scale,  the local flow around the particle can be considered to be inertia-free, and Stokes flow solutions can be used to relate the rotational dynamics of the particles to the local velocity gradient tensor. This problem was considered in the classic paper by \cite{Jeffery22}, who solved the problem of Stokes flow around a general triaxial ellipsoidal object.  {He then
derived, for the special case of an axisymmetric ellipsoid, the evolution equation for the orientation vector} as function of the local velocity gradient tensor. Such dynamics lead to fascinating phenomena such as a rotation of rods when placed in a constant shear (Couette) flow and periodic motions on closed (Jeffery's) orbits. Effects of such motions on the rheology of suspensions has been studied extensively, see e.g. \cite{Larson99}. 

In turbulent flows the velocity gradient tensor $A_{ij} = \partial u_i/\partial x_j$ fluctuates and is dominated by small-scale motions, on the order of the Kolmogorov scale $\eta_K$, and much work has focused on rod-like particles  whose size is smaller than  $\eta_K$. Studies of the orientation dynamics of such particles in turbulent flows have included those of \cite{ShinKoch05,PumirWilkinson11} using isotropic turbulence data from direct numerical simulation (DNS), those of \cite{Zhangetal01} and \cite{Mortensenetal08} for particles in channel flow turbulence also using DNS, and those of \cite{BernsteinShapiro94} and \cite{NewsomBruce98} using data from laboratory and atmospheric measurements, respectively. {(We remark that  \cite{ShinKoch05} also consider fibers that are longer than $\eta_K$).} In numerical studies, Lagrangian tracking is most often used to determine the particle trajectories and simultaneous time integration of the Jeffery equation along the trajectory leads to predictions of the particles' orientation dynamics. 

Generic properties of the orientation dynamics, such as the variance of the fluctuating orientation vector or its alignment trends may also be studied by making certain assumptions about the Lagrangian evolution of the carrier fluid's velocity gradient, in particular about its symmetric and antisymmetric parts, the strain-rate tensor ${\bf S} = \frac{1}{2}({\bf A}+{\bf A}^\top)$ and rotation-rate tensor $\vct{\Omega} = \frac{1}{2}({\bf A}-{\bf A}^\top)$. A number of recent theoretical studies have been based on the assumption that these flow variables obey isotropic Gaussian statistics, e.g. are the result of linear Ornstein-Uhlenbeck processes (see e.g.  \cite{BrunkKoch98,PumirWilkinson11,WilkinsonKennard12,Vincenzi13}). This assumption facilitates a number of theoretical results that may be used to gain insights into some features of the orientational dynamics, such as in the limiting case of strong vorticity with a weak random straining background, in which analytical solutions for the full probability density are possible \citep{Vincenzi13}. {In these studies,  the crucial role of alignments between the particles and the vorticity has been highlighted.
As will be shown in the following, the relative alignment of the vorticity with the strain rate eigendirections, as first observed in \cite{Ashurstetal87}  is also crucially important.}  

In a recent study based on DNS of isotropic turbulence, \cite{Parsaetal12} analyze the orientational dynamics of axisymmetric ellipsoids of any aspect ratio, that is, from rod-like shapes to spherical and disc-like shapes.  They consider axisymmetric ellipsoids with major semi-axes of length $d_1$, $d_2$, $d_3$, with $d_2=d_3$. The unit orientation vector ${\bf p}$ is taken to  point in the direction of the axis of size $d_1$. The parameter $\alpha = d_1/d_2=d_1/d_3$ describes uniquely the type of anisotropy: For $\alpha \to \infty$ one has rod or fiber-like particles with ${\bf p}$ aligned with the axis, while for $\alpha \to 0$ one has discs with ${\bf p}$ aligned perpendicular to the plane of the disc. For $\alpha \to 1$, one has spheres for which the choice of ${\bf p}$ is arbitrary relative to the object's geometry. \cite{Parsaetal12}  report the variance and flatness factors of the orientation vector's rate of variation $ \dot{\bf p}$ in time along fluid tracer trajectories. Strong dependencies of the variance as function of $\alpha$ are observed. The trends differ significantly from results obtained when one assumes that ${\bf p}$ and ${\bf A}$ are uncorrelated, or that ${\bf A}$ follows Gaussian statistics with no preferred vorticity-strain rate alignments.  

Both \cite{ShinKoch05} and \cite{Parsaetal12} observe that  the non-trivial dependencies of the particle rotation variance as function of $\alpha$ are associated with the alignment trends between flow vorticity and strain-rate eigenvectors. They remark that the orientation dynamics of anisotropic particles can thus serve as a useful diagnostic to examine the accuracy of Lagrangian models for the velocity gradient tensor in turbulence.  Several Lagrangian stochastic models for the velocity gradient tensor in turbulence have been proposed in the literature \citep{GirimajiPope90,Cantwell92,Jeonggirimaji03,Chertkovetal99,ChevillardMeneveau06,Nasoetal07,Biferaleetal07}. As reviewed in \cite{Meneveau11}, some of these models are for coarse-grained velocity gradients \citep{Biferaleetal07} or tetrads of fluid particles \citep{Chertkovetal99,Nasoetal07}, while others describe transient or quasi-steady state behavior only \citep{Cantwell92,Jeonggirimaji03}. {To our knowledge, only two stochastic processes for the full velocity gradient tensor that includes realistic strain-vorticity correlations lead to stationary statistics. The first is due to \cite{GirimajiPope90}. It enforces by construction that the pseudo-dissipation is a lognormal process, and several additional free parameters must be prescribed. The second process is the RFDA approach \citep{ChevillardMeneveau06} in which a physically-motivated closure is used to model pressure Hessian and viscous effects}. Recent interest has also focused on the fate of particles larger than $\eta_K$ (see e.g. \citet{Zimmermannetal11}). 

Here we shall focus on the case of inertia-free particles smaller than $\eta_K$,  but of a general ellipsoidal shape that is not necessarily  axisymmetric. The aims of the present work are two-fold: firstly, to generalize the results of \cite{Parsaetal12} to the case of generalized ellipsoidal particles, not restricted to the special case of bodies of revolution. For this purpose, we use a generalization of Jeffery's equation {written in a convenient form by \cite{JunkIllner07} that can be considered as a reformulation 
of earlier developments for triaxial ellipsoidal particles by \cite{Jeffery22,Bretherton62,GierszewskiChaffey78,HinLea79,YarGot97} }. The model, summarized in \S \ref{sec-model}, describes the  dynamics of all three orientation vectors pointing in each of the ellipsoid's major axes. The results depend upon two geometric parameters  $d_1/d_2$ and $d_1/d_3$ that are equal for the axisymmetric cases. We aim to measure the variance and flatness of the rates of change of the three orientation vectors as function of these two parameters. Also, geometric features such as the alignments between these orientation vectors and special local flow directions (vorticity, and strain-rate eigen-directions) will be reported. The results from analysis of DNS are presented in \S \ref{sec-DNS}. The observed  relative alignments highlight the importance of correlations among vorticity and strain-rate eigen-directions in determining the particle orientation dynamics. 

A second goal of this study, presented in \S \ref{sec-stochmodels}, is to study the predictions of several models. As also done by \cite{Parsaetal12} for axisymmetric particles, 
we first consider predictions of the variance of particle rotation assuming the particle alignment is uncorrelated from the velocity gradient tensor. We also consider a Gaussian model of the velocity gradient tensor in which vorticity - strain-rate alignments are absent. Then we consider a stochastic model for the velocity gradient tensor \citep{ChevillardMeneveau06,ChevillardMeneveau07} in which pressure and viscous effects are modeled based on the Recent Fluid Deformation Approximation (RFDA).
This model has been shown to yield realistic predictions of stationary statistics for the velocity gradient in turbulence at moderate Reynolds numbers \citep{Chevillardetal08}. While the model has been generalized to passive  scalars \citep{Gonzalez09}, rotating turbulence \citep{Li11} and MHD turbulence \citep{Hateretal11}, the fate of general triaxial-ellipsoidal anisotropic particles acted upon by a velocity gradient tensor that evolves according to the RFDA model has not yet been examined. The model results are compared with those from DNS and conclusions are presented in \S \ref{conclusions}.  

\section{Evolution equations for anisotropic particle orientation}
\label{sec-model}

\begin{figure}
\centering
\epsfig{file=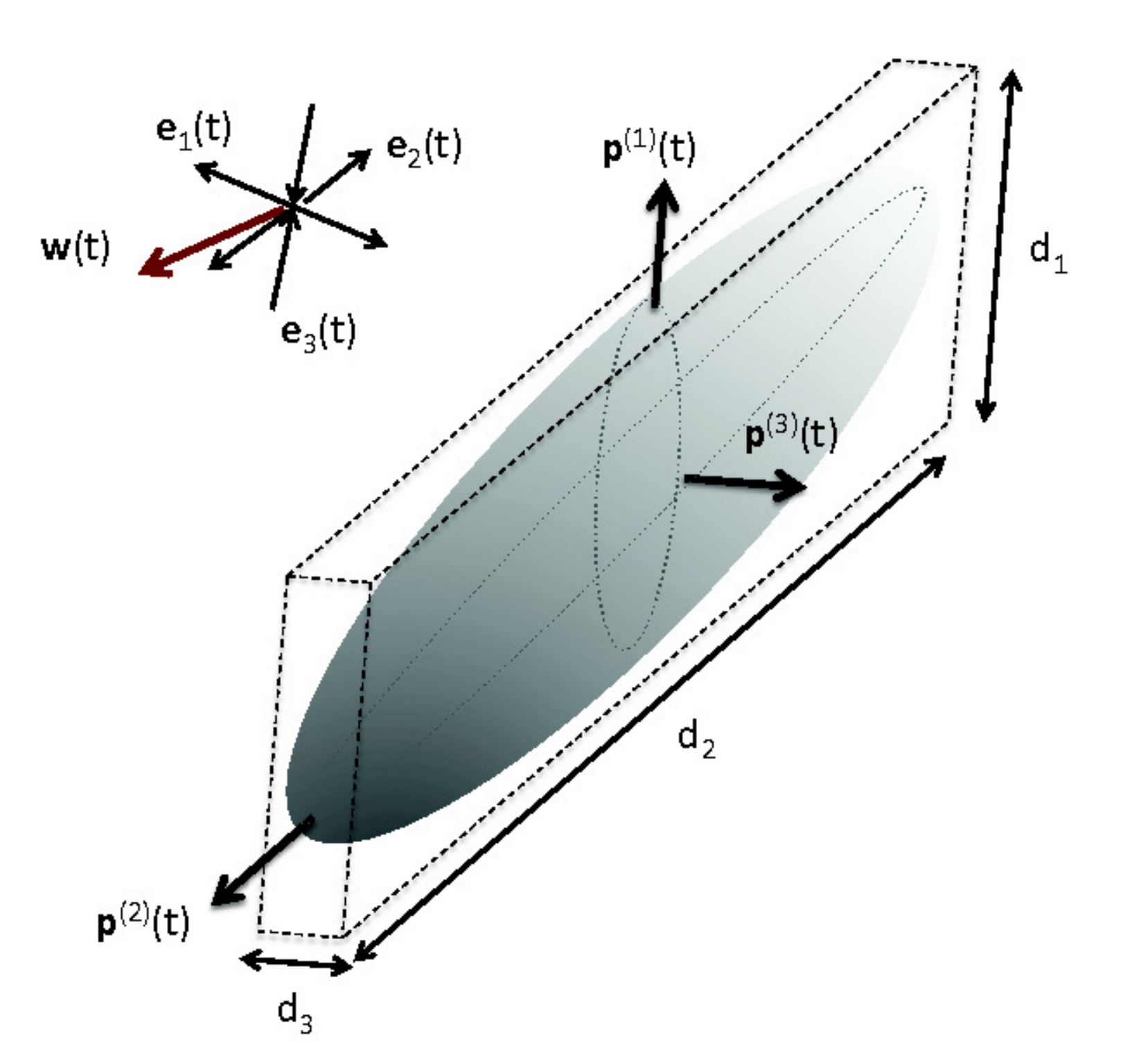,width=7cm}
\caption{\label{fig-sketch} Sketch of triaxial ellipsoidal particle with its three major axes scales $d_1$, $d_2$ and $d_3$, and its respective orientation unit vectors ${\bf p}^{(i)}$ for $i=1,2,3$. Also shown are vorticity $\vct{\omega}$  and strain-rate eigenvectors (most extensional: ${\bf e}_{1}$, intermediate: ${\bf e}_{2}$, and most contracting ${\bf e}_{3}$), characterizing the locally linear but time-dependent flow structure surrounding the particle.}
\end{figure}

Many numerical and theoretical studies use the Jeffery equation \cite{Jeffery22} to predict the time evolution of the orientation of an axisymmetric ellipsoidal particle, as it is advected and acted on by a turbulent velocity field. Specifically, Jeffery's equation  for the unit orientation vector ${\bf p}(t)$ in the ellipsoid major axis of size $d_1$ (while $d_2=d_3$ and $\alpha=d_1/d_2$) reads:
\be
\frac{dp_n}{dt} = \Omega_{nj}p_j+\lambda(S_{nj}p_j-p_np_kS_{kl}p_l),
\label{eq:Jeffery}
\ee
where $\lambda = (\alpha^2-1)/(\alpha^2+1)$ and ${\bf S}$ and $\vct{\Omega}$ are the strain and rotation-rate tensors, respectively. 
This equation is valid for axisymmetric ellipsoids in which two semi-axes are equal.  
The case of more general geometries was considered by \cite{Bretherton62} in which the linearity of Stokes flow and  general
 symmetries gave the general form that any orientation evolution must have.  {Additional references dealing with the orientation dynamics of general ellipsoids include \cite{GierszewskiChaffey78,HinLea79,YarGot97}.  For instance, \cite{HinLea79} found that slight deviations from axisymmetry may cause significant variations in the resulting rates of particle rotation.  In a recent paper \citep{JunkIllner07}, the results were cast in a practically useful form}, namely a system of 3 equations describing the evolution of three perpendicular unit vectors, each directed in one of the principal directions of the  ellipsoid.  Figure \ref{fig-sketch} illustrates the geometry of a general triaxial ellipsoid, exposed to  the action of a surrounding local vorticity and strain rate  field at much larger scales. Since Stokes flow is assumed, the result is applicable only when the size of the tracer particle, i.e. its largest dimension, is smaller than the Kolmogorov scale of turbulence.  

The Junk \& Illner equation reads, for the three perpendicular unit vectors ${\bf p}^{(i)}$ ($i=1,2,3$):
\be
\frac{d{\bf p}^{(i)}}{dt} = \frac{1}{2} (\vct{\nabla} \times {\bf u}) \times {\bf p}^{(i)} + \sum \limits_{k,m} \epsilon_{ikm} \lambda^{(m)} {\bf p}^{(k)} \otimes {\bf p}^{(k)}~ {\bf S}  ~ {\bf p}^{(i)},
\label{eq:JunkIllner}
\ee
where $\otimes$ stands for the tensorial product, i.e. for any vectors $\textbf{u}$ and $\textbf{v}$, we have $(\textbf{u}\otimes\textbf{v})_{ij} = u_iv_j$. Moreover, we sum  over repeated subscript indices (i.e. Einstein convention), but we do not sum over repeated superscripts in parenthesis, unless explicitly stated.  The $\lambda^{(m)}$'s are given by
\be\label{eq:deflambdais}
\lambda^{(1)} = \frac{\left({d_2}/{d_3}\right)^2-1}{\left({d_2}/{d_3}\right)^2+1},~~~~\lambda^{(2)} = - \frac{\left({d_1}/{d_3}\right)^2-1}{\left({d_1}/{d_3}\right)^2+1},~~~~\lambda^{(3)} = \frac{\left({d_1}/{d_2}\right)^2-1}{\left({d_1}/{d_2}\right)^2+1}.
\ee 
The $\lambda^{(m)}$'s are not independent.  Solving (e.g.) for $d_2/d_3$ as function of $\lambda^{(1)}$, one obtains $\lambda^{(1)} = 1/\lambda^{(2)}+1/\lambda^{(3)}$.
Then,  in full index notation, the evolution equation reads:
\be\label{eq:JunkIllnerIndex}
\frac{dp^{(i)}_n}{dt} = \Omega_{nj}p^{(i)}_j+\sum \limits_{k,m} \epsilon_{ikm} \lambda^{(m)} p^{(k)}_n p^{(k)}_q S_{ql} p^{(i)}_l.
\ee
The norm of ${\bf p}^{(i)}$ remains unity, since the inner product of the RHS of Eq. \ref{eq:JunkIllner} by  ${\bf p}^{(i)}$, or equivalently a contraction of the RHS of Eq. \ref{eq:JunkIllnerIndex} with $p^{(i)}_n$, gives a contribution proportional to $p^{(i)}_n p^{(k)}_n = \delta_{ik}$. The $\sum \limits_{k,m} \epsilon_{ikm}$ operation then makes this term vanish. We recall that for any vector $\textbf{u}$, the quadratic form $\textbf{u}^\top {\bf \Omega}\textbf{u}$ vanishes if the tensor ${\bf \Omega}$ is antisymetric.  Also, \cite{JunkIllner07} show that for the case of $d_2=d_3$, i.e. $\lambda^{(1)}=0$ and $\lambda^{(2)}=-\lambda^{(3)} = -\lambda$, one recovers the Jeffery equation:
Consider $i=1$ as the main axis. Then we have 
\begin{subeqnarray}
 \sum \limits_{k,m} \epsilon_{1km} \lambda^{(m)} {\bf p}^{(k)}\otimes {\bf p}^{(k)}~ {\bf S}~ {\bf p}^{(i)}   & = & -\lambda^{(2)} {\bf p}^{(3)} \otimes{\bf p}^{(3)} ~ {\bf S}  ~{\bf p}^{(1)} +\lambda^{(3)} {\bf p}^{(2)}\otimes {\bf p}^{(2)} ~ {\bf S}~  {\bf p}^{(1)}\nonumber \\
 & = & \lambda\left({\bf p}^{(3)}\otimes {\bf p}^{(3)}+ {\bf p}^{(2)}\otimes {\bf p}^{(2)} \right)~ {\bf S}~  {\bf p}^{(1)}.\nonumber
\end{subeqnarray}
Since the three unit vectors form an orthogonal basis, we can use the fact that \[
{\bf p}^{(1)}\otimes {\bf p}^{(1)} +{\bf p}^{(2)}\otimes {\bf p}^{(2)} +{\bf p}^{(3)}\otimes {\bf p}^{(3)} = {\bf I},
\] 
to obtain
\be
\sum \limits_{k,m} \epsilon_{1km} \lambda^{(m)} {\bf p}^{(k)}\otimes {\bf p}^{(k)} ~ {\bf S} ~ {\bf p}^{(i)} =   \lambda\left({\bf I} - {\bf p}^{(1)}\otimes {\bf p}^{(1)} \right)~ {\bf S} ~ {\bf p}^{(1)},
\ee
which is the last term in the Jeffery equation (Eq. \ref{eq:Jeffery}) taking $\textbf{p}^{(1)} =\textbf{p}$.

Under the assumption of small inertia-free particles,  the ellipsoid's centroid follows the fluid flow. Therefore,  the time evolution in Eq. \ref{eq:JunkIllner} can be interpreted as the evolution along fluid particle trajectories and the time evolution of the orientations depends solely on the velocity gradient tensor. Knowing $\Omega_{ij}(t)$ and $S_{ij}(t)$ along fluid trajectories thus allows us to evaluate the Lagrangian evolution of particle orientations ${\bf p}^{(i)}$ by solving Eq. \ref{eq:JunkIllner} 
(one may of course only solve for two components, since at all times, e.g., $\textbf{p}^{(3)}=\textbf{p}^{(1)} \times \textbf{p}^{(2)}$).  

In this article, we are interested in the variance of the rate of rotation of these orientation vectors (tumbling rates), extending the results for the variance of $\textbf{p}^{(1)}$ when $d_2=d_3$ and $d_1/d_2\rightarrow +\infty$ (i.e. rods) studied in \cite{ShinKoch05}, and those of  \cite{Parsaetal12} for any $d_1/d_2$ with $d_2=d_3$ (i.e. from rods to discs). Specifically, we are interested in the variance of the rotation speed of each direction vector as function of the two independent anisotropy parameters, 
\begin{equation}\label{eq:Varpi}
V^{(i)}\left(\frac{d_1}{d_2},\frac{d_1}{d_3}\right)=\frac{1}{2\langle \Omega_{rs}\Omega_{rs}\rangle}\langle \dot{p}_n^{(i)}  \dot{p}_n^{(i)} \rangle
\end{equation}
for the three orientation vectors $i=1,2,3$, and we recall that no summation is assumed over superscripts  $(i)$. In Eq. \ref{eq:Varpi}, the average of the square norm $|\dot{\textbf{p}}^{(i)}|^2$ is normalized by the average rate-of-rotation of the flow. We recall that for isotropic turbulence, $2\langle \Omega_{rs}\Omega_{rs}\rangle =2\langle S_{rs}S_{rs}\rangle = \varepsilon/\nu$, where $\varepsilon$ is the average dissipation per mass and $\nu$ the kinematic viscosity. Following \cite{Parsaetal12} we are also interested in the flatness factor, namely
\begin{equation}\label{eq:Flatpi}
F^{(i)}\left(\frac{d_1}{d_2},\frac{d_1}{d_3}\right)=\frac{\langle(\dot{p}_n^{(i)} \dot{p}_n^{(i)})^2\rangle}{\langle \dot{p}_r^{(i)} \dot{p}_r^{(i)}\rangle^2}
\end{equation} 
Further extending the prior analyses, we are also interested in the alignments between ${\bf p}^{(i)}$ and the vorticity and strain-rate eigenvectors in the flow. 

\section{Orientation dynamics of triaxial ellipsoids in DNS of isotropic turbulence}
\label{sec-DNS}

In this section, we consider orientation dynamics of triaxial ellipsoids in isotropic turbulence at moderate Reynolds number. We use results from Direct Numerical Simulation (DNS) of forced isotropic turbulence at $R_{\lambda} \approx 125$ to provide Lagrangian time-histories of the velocity gradient tensor $A_{ij}(t)$ along the tracer particle trajectories. These data are the same as those used in \cite{ChevillardMeneveau11}. The DNS is based on a pseudo-spectral method, de-aliased according to the $\frac{3}{2}$-rule and with 2nd-order accurate Adams-Bashforth time stepping. The computational box is cubic (size $2\pi$) with periodic boundary conditions in the three
directions and a spatial mesh with $256^3$ grid points. Statistical stationarity is maintained by an isotropic external force acting
at low wavenumbers in order to ensure a constant  power injection. It provides, in the units of the simulation,
a constant energy injection rate of $\epsilon= 0.001$. The kinematic viscosity of the fluid is $\nu = 0.0004$. The Kolmogorov
scale is $\eta_K =  0.016$ so that $\Delta x/\eta_K \approx 1.5$ (with $ \Delta x = 2\pi/256$). Lagrangian trajectories are obtained using 
numerical integration of the fluid tracer equation. A  second order time integration scheme is used (involving both velocity and acceleration at the current spatial location), with a time step $\Delta t = 3.5 ~10^{-3}$  and cubic spatial interpolation in the three spatial directions to obtain velocity and acceleration at particle locations.  A total of 512 such trajectories, of length of order of 15 large-eddy turnover times, is used in this study, with initial positions chosen at random in the
flow volume. {We have tested for statistical convergence levels using fewer and shorter trajectories (data not shown) and obtained results with negligible deviations from those plotted. Statistical quantities shown, including the flatness, have converged statistically to levels equal to or smaller than the amplitudes of the deviations from smooth behavior shown in the
plots (i.e. better than about 3\%)}.  Once we have the Lagrangian time history of the velocity gradient tensor $\textbf{A}(t)$, we integrate numerically the dynamics of the three vectors $\textbf{p}^{(i)}$ {starting from three unit vectors that point in arbitrary (randomly chosen) initial directions, constrained by the orthonormality condition ${p_q}^{(i)}(0) = \delta_{iq}$}. 
For high accuracy (e.g. always checking that the set of three vectors remains orthonormal), we integrate Eq. \ref{eq:JunkIllner} using a standard fourth-order Runge-Kutta method, using the same time-step as used in the time integration to obtain the time history of $\textbf{A}(t)$.

\subsection{Variance and Flatness of the time derivative of orientation vectors}
\begin{figure}
\centering
\epsfig{file=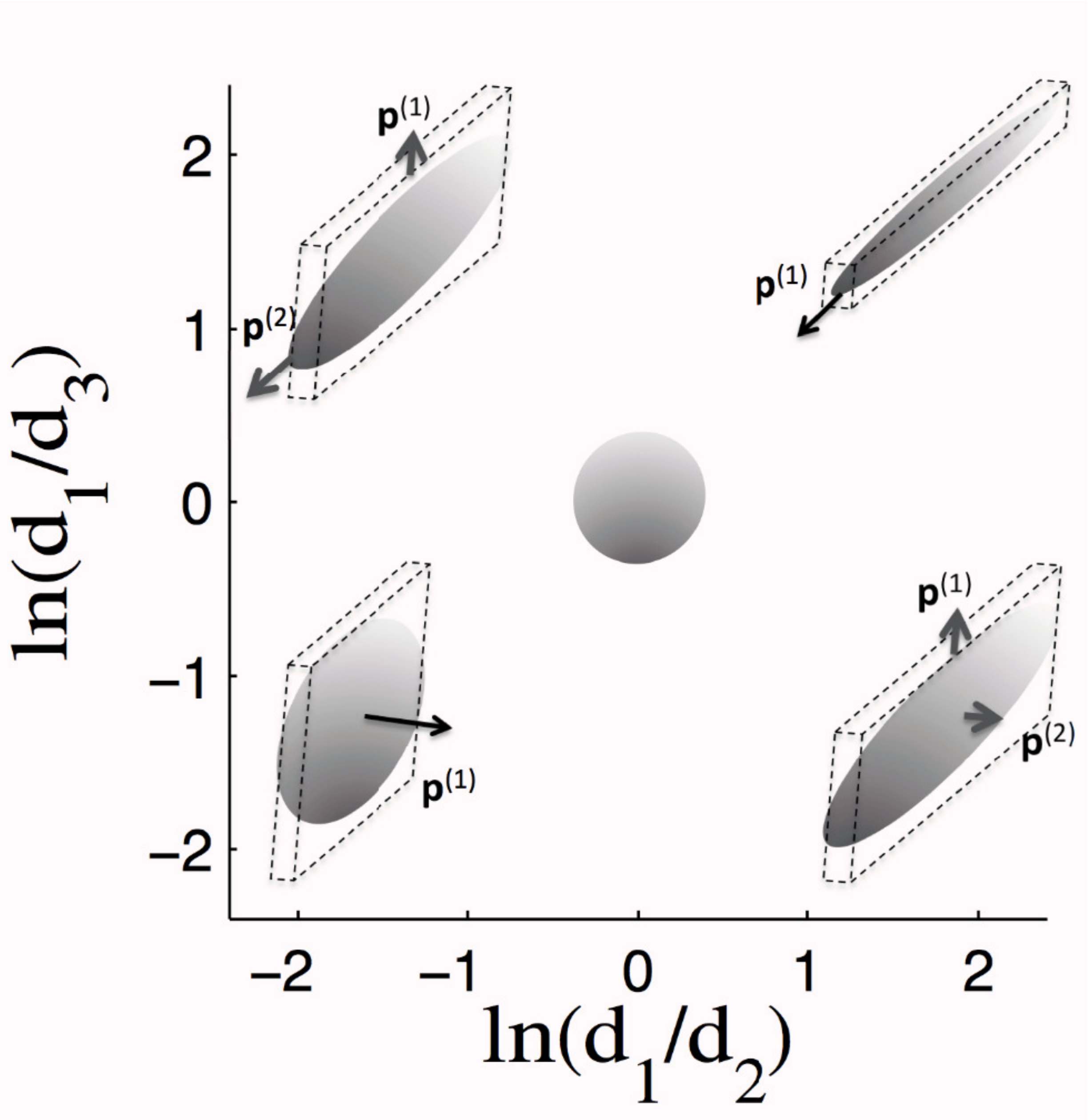,width=8cm}
\caption{\label{fig-sketchquadrants}  
Sketch of triaxial ellipsoid geometries and orientation vectors, as function of semi-axes ratios $d_1/d_2$ and $d_1/d_3$.}
\end{figure}

We focus in this section on the variance $V^{(i)}\left(\frac{d_1}{d_2},\frac{d_1}{d_3}\right)$ of the orientation vectors as defined in Eq. \ref{eq:Varpi}. The sketch in Fig. \ref{fig-sketchquadrants} represents the various limiting geometries and orientation vectors as function of the ratios $d_1/d_2$ and $d_1/d_3$.
Due to the symmetries inherent in the labeling of the various directions, interchanging $d_2$ and $d_3$ while leaving $d_1$ unchanged should not affect the rotation rates in the $i=1$ direction. 
Hence,  the results are expected to be symmetric around the 45-degree line ($d_1/d_2=d_1/d_3$), i.e.  
\be V^{(1)}\left(\frac{d_1}{d_2},\frac{d_1}{d_3}\right) = V^{(1)}\left(\frac{d_1}{d_3},\frac{d_1}{d_2}\right).
\ee 
Also, since upon interchanging the two directions (2) and (3), the corresponding direction vector must be interchanged, we expect 
\be V^{(3)}\left(\frac{d_1}{d_3},\frac{d_1}{d_2}\right)=V^{(2)}\left(\frac{d_1}{d_2},\frac{d_1}{d_3}\right),
\ee
and hence only results for $V^{(1)}$ and $V^{(2)}$ are presented. The same symmetries apply to the flatness factor,  and hence only results for $F^{(1)}$ and $F^{(2)}$ are presented. 

\begin{figure}
\centering
\epsfig{file=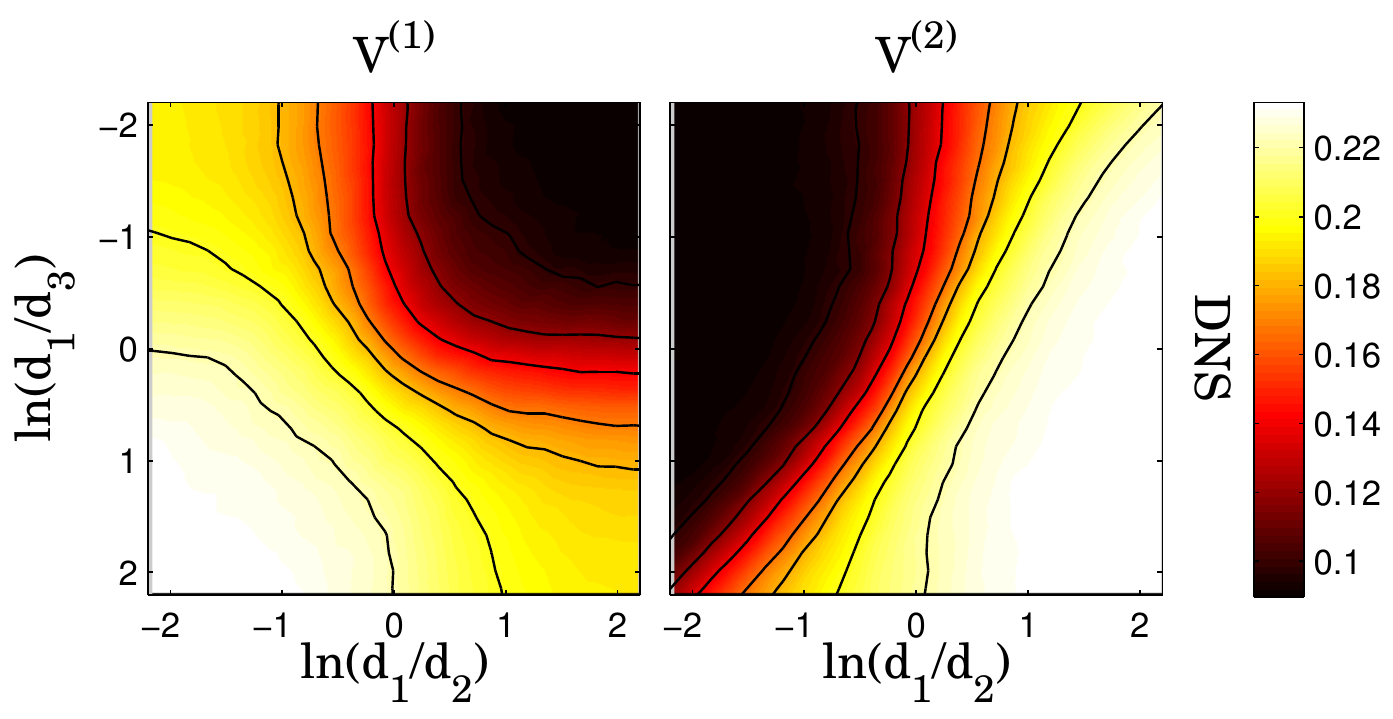,width=11cm}
\caption{\label{fig-variance-p-DNS}  Variance of rate-of-change of two ellipsoid orientation vectors ${\bf p}^{1}$ and ${\bf p}^{2}$ as function of the two ratios of semi-axes length, obtained from DNS. Contour lines go from 0.1 to 0.22 separated by 0.02.}
\end{figure}

In figure \ref{fig-variance-p-DNS} we show the normalized variance as function of the two ratios of semi-axes length, obtained from DNS.
Several observations may be made based on these results. Firstly, for axisymmetric ellipsoids along the $d_1/d_3=d_1/d_2$ line,
the results for $V^{(1)}$  agree with those of \cite{Parsaetal12}. Namely, for fiber-like ellipsoids ($d_1/d_2 \to \infty$), the normalized variance tends to 
values near 0.09, whereas for disc-like ellipsoids, it tends to values near 0.24. For spherical particles, it is near 0.17. 
The trend at the edges of the negative 45 degree line, $d_1/d_3=(d_1/d_2)^{-1}$ represent particles that are long in one direction (e.g. $d_2\gg d_1$), very thin in another ($d_3\ll d_1$), and of  intermediate size ($d_1$) in the direction chosen for ${\bf p}^{(1)}$. As can be seen, $V^{(1)}$ along this line remains near the spherical value, with a small increase towards $V^{(1)}(d_1/d_2 \to 0, d_1/d_3 \to \infty) \sim 0.19$. 

The variance $V^{(2)}$ of the orientation vector in a direction perpendicular to ${\bf p}^{(1)}$ and along the direction of either the longest or shortest ellipsoid axis exhibits significant dependence upon the semi-axes scale ratios. For ${\bf p}^{(2)}$ aligned along the largest ellipsoid semi-axis (top-left corner of the figure), the variance is reduced, to about 0.09. This is similar to the variance for long axisymmetric fibers. For ${\bf p}^{(2)}$ aligned along the shortest ellipsoid semi-axis (bottom-right corner of the figure), the variance is large, on the order of 0.24, similar to the values for axisymmetric discs. 
It was noted by \cite{Parsaetal12} that the transition between rod and disc-like behaviors occurred quite rapidly, with aspect ratios of about $d_1/d_2 \sim $
2-3 already showing results quite close to the asymptotic values. As can be seen in the results for $V^{(2)}$, the transition is
even more rapid along the negative 45 degree, $d_1/d_3=(d_1/d_2)^{-1}$ line, where most of the change in variance occurs for values between $d_1/d_2 \sim$  0.6 and 1.6. 

\begin{figure}
\centering
\epsfig{file=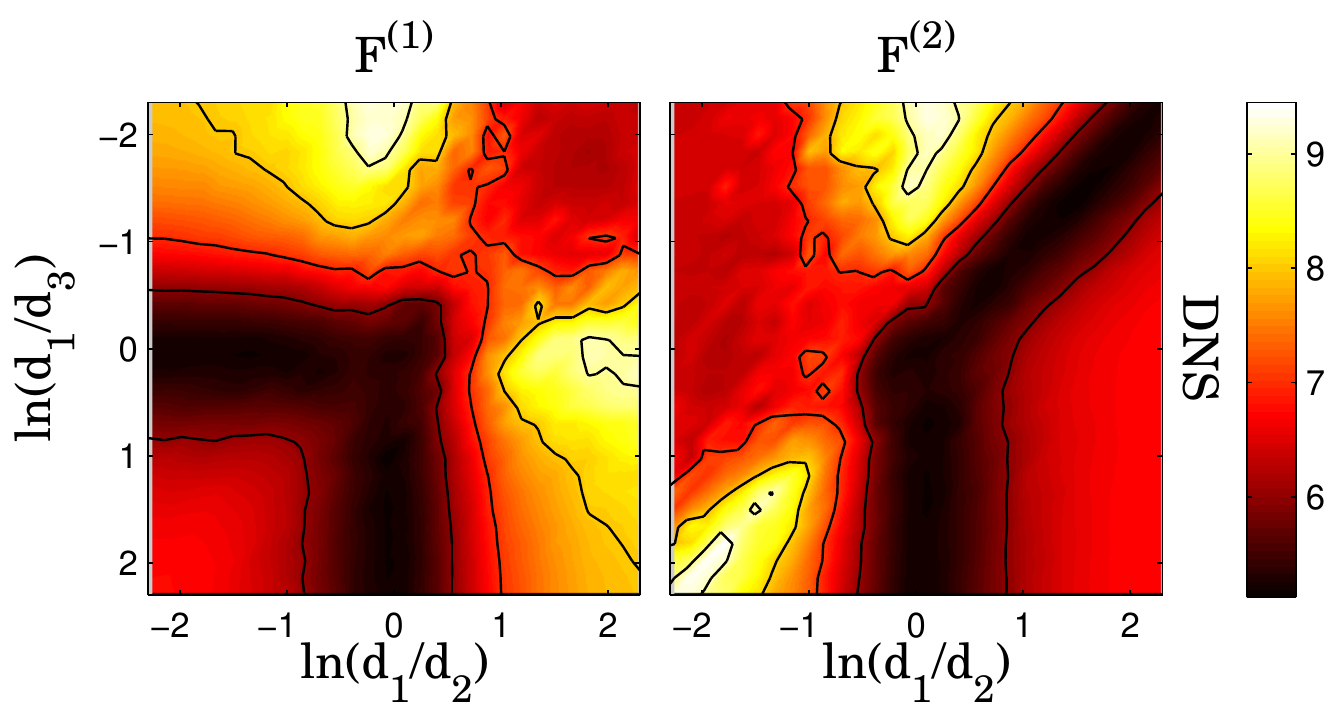,width=11cm}
\caption{\label{fig-flatness-p-DNS}  Flatness of rate-of-change of two ellipsoid orientation  vectors ${\bf p}^{(1)}$ and  ${\bf p}^{(2)}$ as function of the two ratios of semi-axes length, obtained from DNS. Contour lines correspond to Flatness values of 2.1, 2.2, 3, 3.8, 5, 6, 7, 8, 9.}
\end{figure}

Next, the flatness factors of the orientation rates-of-change, $F^{(1)}$ and $F^{(2)}$ (Eq. \ref{eq:Flatpi}) are presented, in Fig. \ref{fig-flatness-p-DNS}. As found by \cite{Parsaetal12} for axisymmetric cases, the flatness is in a range between 5 and 10. These values are clearly above $5/3$, which is the value obtained when the vector $\dot{\bf p}^{(1)}$ is assumed to have zero-average independent Gaussian components  \citep{Parsaetal12} or 2, which is the value obtained for spheres (i.e. $d_1=d_2=d_3$) when $\dot{\bf p}^{(1)}$ is assumed independent from velocity gradients, themselves assumed Gaussian (see Appendix and Eq. \ref{eq:FlatSIspheres}). 
The maximum flatness is observed near the top middle and right middle regions, where $d_1 \approx d_2$ and $d_3 \ll d_1$ or 
$d_1 \approx d_3$ and $d_1\approx d_2$, respectively, i.e. disc-like shapes, but with ${\bf p}^{(1)}$ aligned in the plane of the disc. These were cases where  the variance is
relatively small (see Fig. \ref {fig-variance-p-DNS}).   For  $F^{(2)}$, the structure is more complex, but the limiting cases showing peak flatness values are consistent with the results 
for  $F^{(1)}$: namely the peaks occur for disc-like shapes with the orientation vector aligned in the plane of the disc.  
 {Consistent with the values of flatness that significantly exceed the Gaussian value and equivalent to the results 
of \cite{Parsaetal12}, the probability density functions of $\dot{p}_n^{(i)} \dot{p}_n^{(i)}$ show elongated tails (data not shown).}

\subsection{Alignments of orientation vectors with vorticity and strain eigenframe}
\begin{figure}
\centering
\epsfig{file=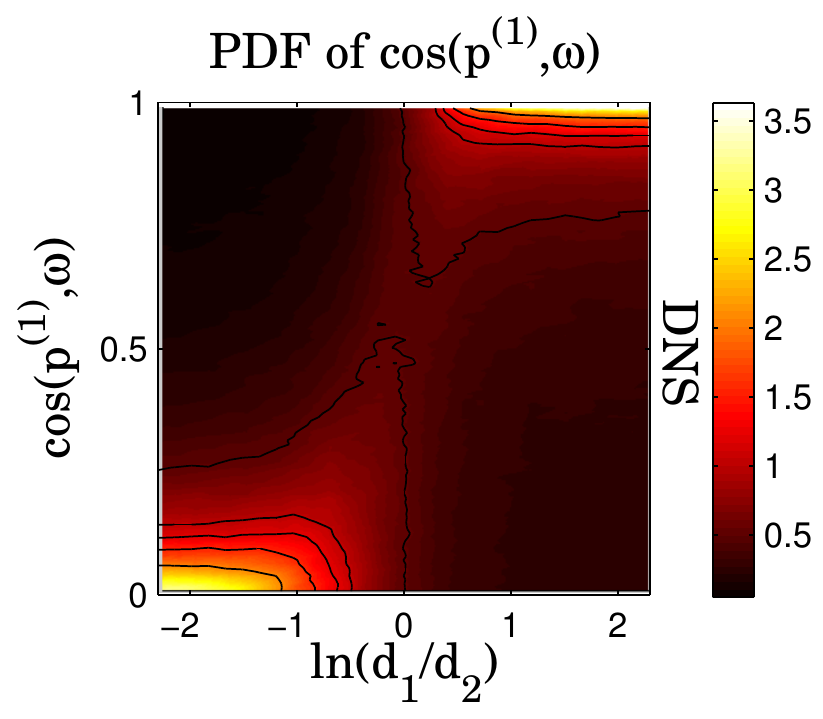,width=7cm}
\caption{\label{fig-align-pw-DNS} Probability-density function (PDF) of cosine of angle between the vorticity direction $\hat{\vct{\omega}}$ and the ellipsoid's major axis ${\bf p}^{(1)}$ for axisymmetric case (i.e. $d_2=d_3$), as function of the anisotropy  parameter $d_1/d_2$, obtained from DNS. Contour lines correspond to PDF values of 0.5, 1, 1.2, 1.5, 2.}
\end{figure}

Next, we consider the alignment trends of particle orientation with respect to the vorticity and strain-rate tensor's eigen-directions. For this discussion, we focus on the case of axisymmetric particles and hence focus only on the single orientation vector ${\bf p}^{(1)}$. Alignment trends with vorticity are quantified by measuring the probability-density-function (PDF)  of $\cos(\theta_{p^1\omega}) = {\bf p}^{(1)} \cdot \hat{\vct{\omega}}$ of the angle between $ {\bf p}^{(1)}$ and the vorticity direction $\hat{\vct{\omega}} = \vct{\omega}/|\vct{\omega}|$. Results are shown in Fig. \ref{fig-align-pw-DNS} as function of the parameter $\alpha=d_1/d_2$. 

As can be seen, for fibre or rod-like particles ($d_1/d_2 \to \infty$), the results confirm strong alignment with vorticity, a well-known trend found in many 
prior studies of alignments of line elements in turbulence \citep{ShinKoch05,Girimajipope90b,PumirWilkinson11}. In the other limit, for
disc-like particles, the results show that  ${\bf p}^{(1)}$ is more preferentially  perpendicular to the vorticity. That is to say, the vorticity tends to be in the 
plane of the disc. These orientation trends help understand the parameter dependencies seen in the variance of 
$\dot{\textbf{p}}^{(1)}$, i.e. $V^{(1)}$ shown in Fig. \ref{fig-variance-p-DNS}. Specifically,  along the diagonal $d_1/d_2=d_1/d_3$ for long rods (i.e. $d_1/d_2\to \infty$), the particle rotates along its axis of symmetry since $\textbf{p}^{(1)}$ is preferentially aligned with the vorticity. This leads to a reduced level of fluctuations $V^{(1)}$ \citep{ShinKoch05,Parsaetal12}. For discs, i.e. $d_1/d_2\to 0$, the vorticity is preferentially aligned  in the plane of the disc which, like a spinning coin on a table, implies faster rotation of the orientation vector perpendicular to that plane. 
 
\begin{figure}
\centering
\epsfig{file=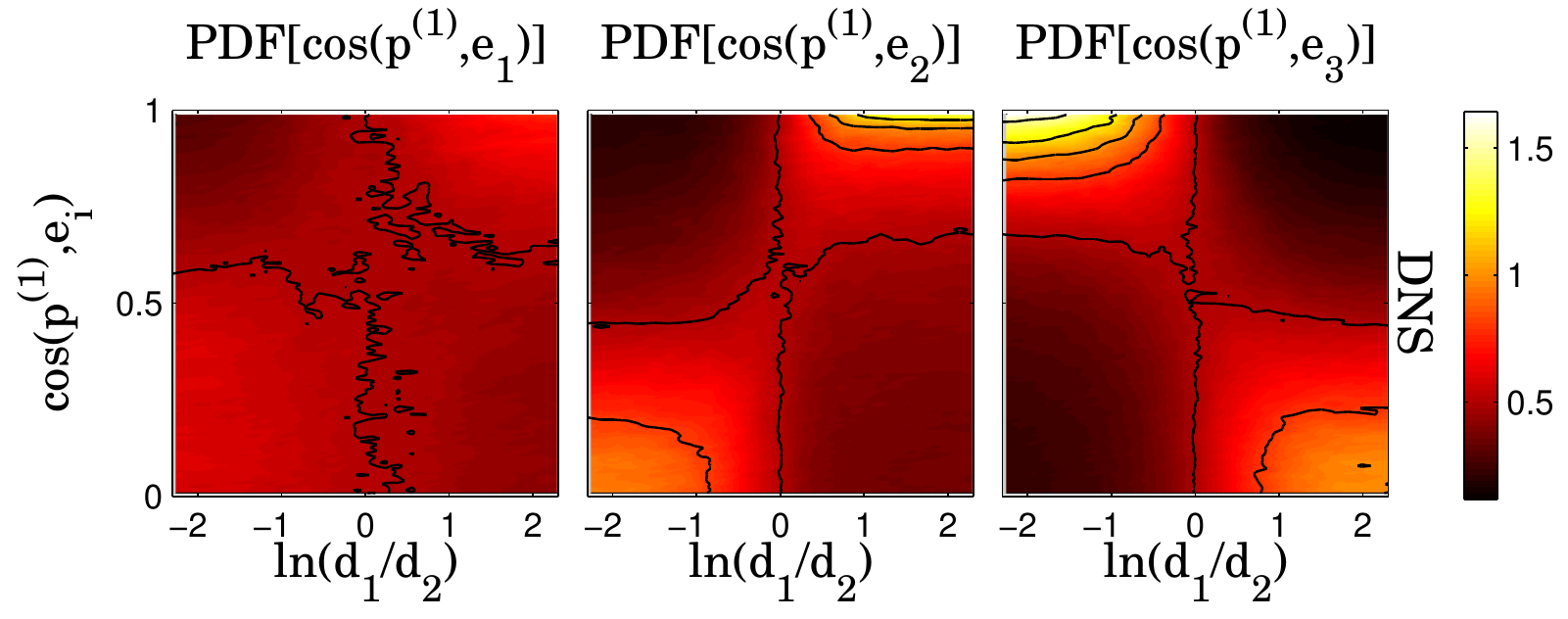,width=13.5cm}
\caption{\label{fig-align-pe-DNS}  PDF of cosine of angle between the strain-rate tensor eigen-directions ${\bf e}_{i}$ and the axisymmetric ellipsoid's major axis ${\bf p}^{(1)}$ for axisymmetric case, as function of the anisotropy parameter $d_1/d_2=d_1/d_3$, obtained from DNS. ${\bf e}_{1}$ is the direction of strongest extension (i.e. most postive eigenvalue), whereas ${\bf e}_{3}$ is the direction of strongest contraction (i.e. most negative eigenvalue). Contour lines correspond to PDF values of 0.5, 0.8, 1, 1.2, 1.5.}
\end{figure}

A similar analysis is done for alignments with each of the strain-rate eigen-directions. The alignments of ${\bf p}^{(1)}$ with each of the strain-rate eigenvectors are quantified using the PDF of the respective angle cosines. The results results are shown in Fig. \ref{fig-align-pe-DNS}.  The strongest alignment trend observed from the DNS is for disc-like particles to align with the most contracting eigen-direction (right panel of Fig. \ref{fig-align-pe-DNS}) . This trend is quite easy to understand intuitively: 
the disc-plane tends to become perpendicular to the incoming (contracting) relative local flow direction. 
And, recall that the vorticity is perpendicular to the most contracting direction (data not shown, see \cite{Meneveau11}). 
The other trend that is visible is that rod-like particles tend to become perpendicular to the most contracting direction. This trend is consistent with its alignment with the vorticity, which tends to be perpendicular to the most contractive direction. As previously observed, rod-like particles tend to align well with the intermediate eigen-vectors, while disc-like particles show preponderance of perpendicular orientation with regards to the intermediate eigen-vector.  Interestingly, alignment with the most stretching eigen-direction appears to be very weak, almost random. 

\begin{figure}
\centering
\epsfig{file=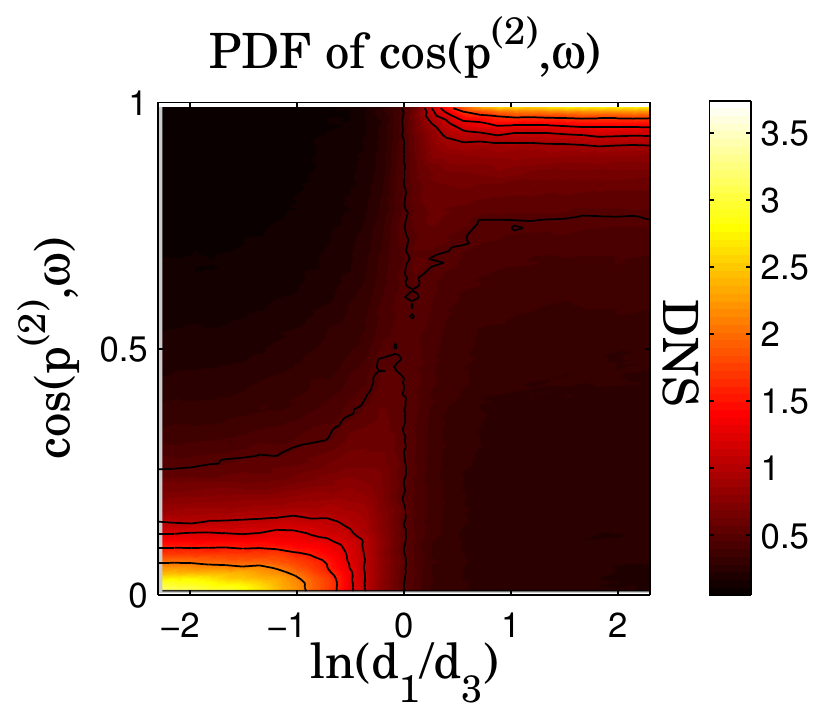,width=7cm}
\caption{\label{fig-align-p2w-DNS}  Probability-density of cosine of angle between the vorticity direction $\hat{\vct{\omega}}$ and the ellipsoid's second semi-axis ${\bf p}^{(2)}$, as function of the anisotropy parameters $d_1/d_3 = (d_1/d_2)^{-1}$, obtained from DNS. Contour lines correspond to PDF values of 0.5, 1, 1.2, 1.5, 2.}
\end{figure}

\begin{figure}
\centering
\epsfig{file=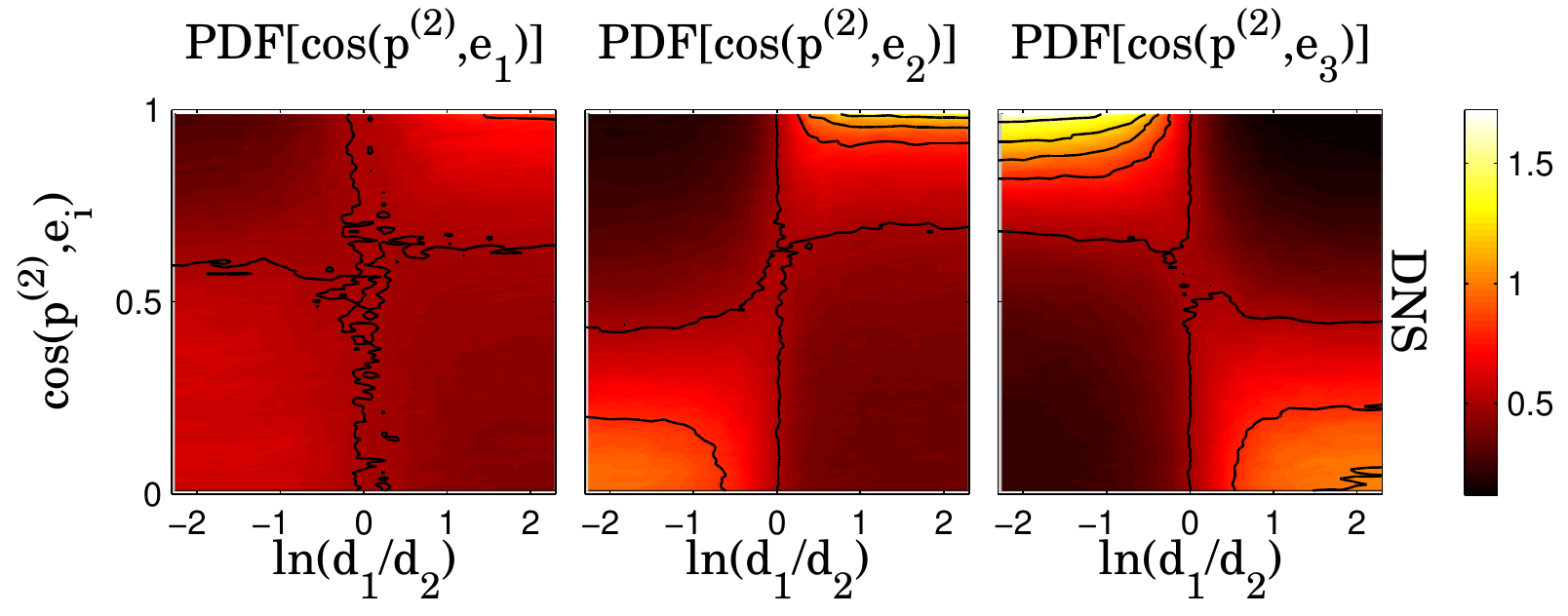,width=13.5cm}
\caption{\label{fig-align-p2e-DNS}  Probability-density of cosine of angle between the strain-rate tensor eigen-directions ${\bf e}_{i}$ and the ellipsoid's second semi-axis ${\bf p}^{(2)}$, as function of the anisotropy  parameter $d_1/d_2$, choosing $d_1/d_3 = (d_1/d_2)^{-1}$, obtained from DNS. Contour lines correspond to PDF values of 0.5, 0.8, 1, 1.2, 1.5.}
\end{figure}

In order to examine alignment trends in non-axisymmetric cases, we consider  the case
along the negative 45 degree diagonal  $d_1/d_3 = (d_1/d_2)^{-1}$. According to Fig. \ref{fig-variance-p-DNS} (right panel) along this line there are
very strong variations of the rotation rate variance of ${\bf p}^{(2)}$, specifically with very large variance
when $d_1/d_2 \gg 1$ and  $d_1/d_3 \ll 1$ (bottom right corner). This is the rate of rotation of the vector aligned in the direction of the smallest of the three semi-axes ($d_2$). 
To explain these trends it is of interest to consider the  orientation  statistics of the vector ${\bf p}^{(2)}$ with respect to vorticity and strain-rate eigen-directions, along the $d_1/d_3 = (d_1/d_2)^{-1}$ line. The results are presented in Figs. \ref{fig-align-p2w-DNS} and \ref{fig-align-p2e-DNS}.  Very similar results are obtained as in Figs. \ref{fig-align-pw-DNS} and \ref{fig-align-pe-DNS}. Namely, when fluctuations of rotation rate of ${\bf p}^{(2)}$  are seen to be high, e.g.  near $d_1/d_3 = (d_1/d_2)^{-1}\to 0$,  this is accompanied by ${\bf p}^{(2)}$ being preferentially orthogonal to the vorticity direction. Conversely, when fluctuations are lower, e.g. when $d_1/d_3 = (d_1/d_2)^{-1}\to \infty$,  one sees that ${\bf p}^{(2)}$ and the vorticity are preferentially aligned.

\section{Predictions from stochastic models}
\label{sec-stochmodels}

\subsection{Statistically independent orientations and velocity gradient tensor}
First, we consider the case when the particle orientation vectors ${\bf p}^{(i)}$ and the velocity gradient tensor ${\bf A}$ are assumed to be
statistically independent, as proposed in \cite{ShinKoch05}. A similar analysis has been presented for axisymmetric particles in \cite{Parsaetal12}, leading to the following result for the variance of the rotation rate of the orientation vector:
\begin{equation}\label{eq:VarpiParsa}
\frac{\langle \dot{p}_n  \dot{p}_n \rangle_{\rm SI}}{2\langle \Omega_{rs}\Omega_{rs}\rangle} = \frac{1}{6}+\frac{1}{10}\lambda^2,
\end{equation}
where SI stands for \textit{Statistically Independent} and, as before, the vector ${\bf p}$ coincides with ${\bf p}^{(1)}$, and $\lambda = -\lambda^{(2)}=\lambda^{(3)}$ (see Eq. \ref{eq:deflambdais}).

Generalization of the approach to the case of triaxial ellipsoids involves similar steps (summarized in Appendix A), namely squaring
each side of equation \ref{eq:JunkIllnerIndex}, averaging, and then using the assumed independence between the orientation vectors and the strain-rate and rotation-rate
tensors to separate the averages. Then, isotropic tensor forms are assumed. As shown in Appendix \ref{annex:Calc}, the result is (see Eq. \ref{eq:finalresAnn})
\begin{equation}\label{eq:ExactIndept}
V^{(i)}_{\rm SI} = \frac{1}{6} +\frac{1}{20} \sum_{k\ne i}\left(\lambda^{(k)}\right)^2.
\end{equation}
One recovers the result of \cite{Parsaetal12} (Eq. \ref{eq:VarpiParsa}) using Eq. \ref{eq:ExactIndept} for $d_2=d_3$, i.e.  $-\lambda^{(2)}=\lambda^{(3)}=\lambda $.

\begin{figure}
\centering
\epsfig{file=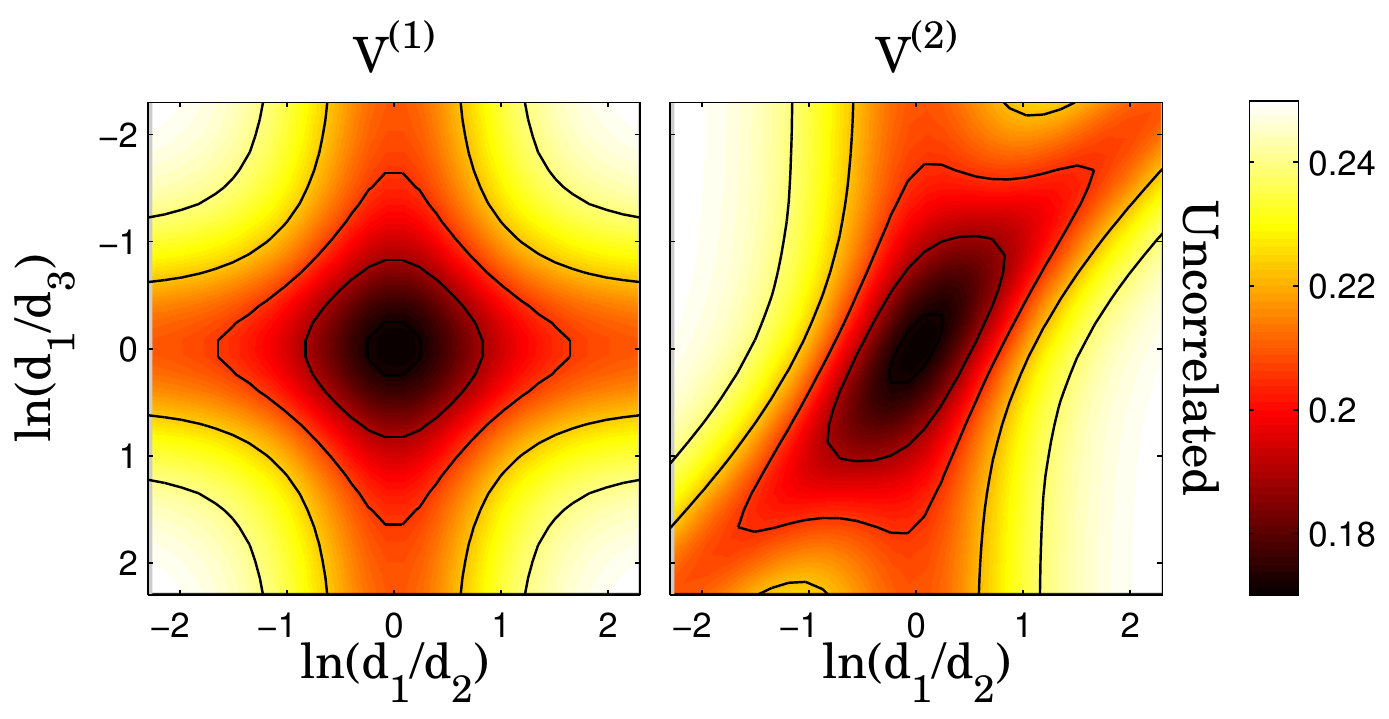,width=11cm}
\caption{\label{fig-variance-p-Uncorrelated}  Variance of rate-of-change of two ellipsoid orientation vectors ${\bf p}^{1}$ and ${\bf p}^{2}$ as function of the two ratios of semi-axes length, when assuming independence of orientation vectors and turbulence velocity gradient tensor elements (from Eq.  \ref{eq:ExactIndept}). Contour lines go from 0.17 to 0.25 separated by 0.02.}
\end{figure}

The results from assuming statistical independence between particle orientation and turbulent velocity gradient tensor are shown in Fig. \ref{fig-variance-p-Uncorrelated}. Comparing $V^{(1)}_{\rm SI}$ with the results of DNS (see Fig. \ref{fig-variance-p-DNS}, left panel), one can see significant differences. As already noted in \cite{ShinKoch05}, the assumption  of independence leads to an over prediction of fluctuations, $V^{(1)}_{\rm SI} \approx 0.26$,  for rods (i.e. top-right corner), instead of $V^{(1)} \approx 0.09$ for DNS. Interestingly, the level of fluctuations predicted for disc-shaped particles (i.e. bottom-left quadrant) is close to the one observed in DNS. Still, the assumption of statistical independence makes no difference between rods and discs. Similar conclusions can be reached about  $V^{(2)}$ (Fig. \ref{fig-variance-p-Uncorrelated}, right panel): DNS leads to marked  differences between the cases $d_1/d_3 = (d_1/d_2)^{-1}\to 0$ and $d_1/d_3 = (d_1/d_2)^{-1}\to \infty$, whereas the  assumption of statistical independence (Eq. \ref{eq:ExactIndept}) does not lead to such differences.

\subsection{Gaussian process}
\label{sec-gaussian}
 
Next, we consider a linear Ornstein-Uhlenbeck process for the velocity gradient tensor according to
\begin{equation}\label{eq:Gaussian}
d{\bf A} =  -\frac{1}{\tau_\eta}{\bf A} dt+\frac{1}{\tau_\eta^{3/2}}d{\bf W}\mbox{ .}
\end{equation}
The term $\bf{W}$ is a tensorial delta-correlated noise that forces the equation. The relaxation term involves a simple time-scale $\tau_\eta$, i.e. the Kolmogorov time scale. In this linear equation, the 1-point covariance structure of the velocity gradients ${\bf A}$ is imposed by the covariance structure of the tensorial forcing term d{\bf W}, whereas the damping term $-\frac{1}{\tau_\eta}{\bf A}$ enforces an exponential time correlation. To ensure isotropic statistics for ${\bf A}$, we use (see Appendix A of \cite{Chevillardetal08})
$$ dW_{ij}(t) = D_{ijpq}dB_{pq}(t),$$  
where $\textbf{B}$ is a tensorial Wiener process with independent elements, i.e. its increments are Gaussian, independent and satify
$$ \langle dB_{pq}\rangle=0 \mbox{ and }\langle dB_{ij}(t)dB_{kl}(t)\rangle=2dt\delta_{ik}\delta_{jl},$$
and the diffusion kernel $D_{ijpq}$ is chosen as
\begin{equation}\label{eq:SolIsotropic}
D_{ijpq}
=\frac{1}{3}\frac{3+\sqrt{15}}{\sqrt{10}+\sqrt{6}}\delta_{ij}\delta_{pq}-\frac{\sqrt{10}+\sqrt{6}}{4}\delta_{ip}\delta_{jq}
+\frac{1}{\sqrt{10}+\sqrt{6}}\delta_{iq}\delta_{jp}\mbox{ .}
\end{equation}
For such a process, one obtains in the stationary regime
$$ \langle A_{ij}(t)A_{kl}(t+\tau) \rangle \build{=}_{t\to \infty}^{}\frac{1}{\tau_\eta^2} e^{-\frac{|\tau|}{\tau_\eta}}\left[ 2\delta_{ik}\delta_{jl}-\frac{1}{2}\delta_{ij}\delta_{kl}-\frac{1}{2}\delta_{il}\delta_{jk}\right],$$
which is consistent with a covariance structure of a trace-free, homogeneous and isotropic tensor, exponentially correlated in time, and such that the variance of diagonal (resp. off-diagonal) elements is $\tau_\eta^{-2}$ (resp. $2\tau_\eta^{-2}$). Accordingly, the covariance structure of its symmetric part is
$$ \langle S_{ij}(t)S_{kl}(t+\tau) \rangle \build{=}_{t\to \infty}^{}\frac{1}{4\tau_\eta^2} e^{-\frac{|\tau|}{\tau_\eta}}\left[ 3\delta_{ik}\delta_{jl}-2\delta_{ij}\delta_{kl}+3\delta_{il}\delta_{jk}\right],$$
and 
$$ \langle \Omega_{ij}(t)\Omega_{kl}(t+\tau) \rangle \build{=}_{t\to \infty}^{}\frac{5}{4\tau_\eta^2} e^{-\frac{|\tau|}{\tau_\eta}}\left[ \delta_{ik}\delta_{jl}-\delta_{il}\delta_{jk}\right]$$
for the anti-symmetric part. Remark that with this definition of $\tau_\eta$, we get $\langle 2S_{pq}S_{pq}\rangle =\langle 2\Omega_{pq}\Omega_{pq}\rangle = \langle A_{pq}A_{pq}\rangle = 15/\tau_\eta^2 = \varepsilon/\nu$.

\begin{figure}
\centering
\epsfig{file=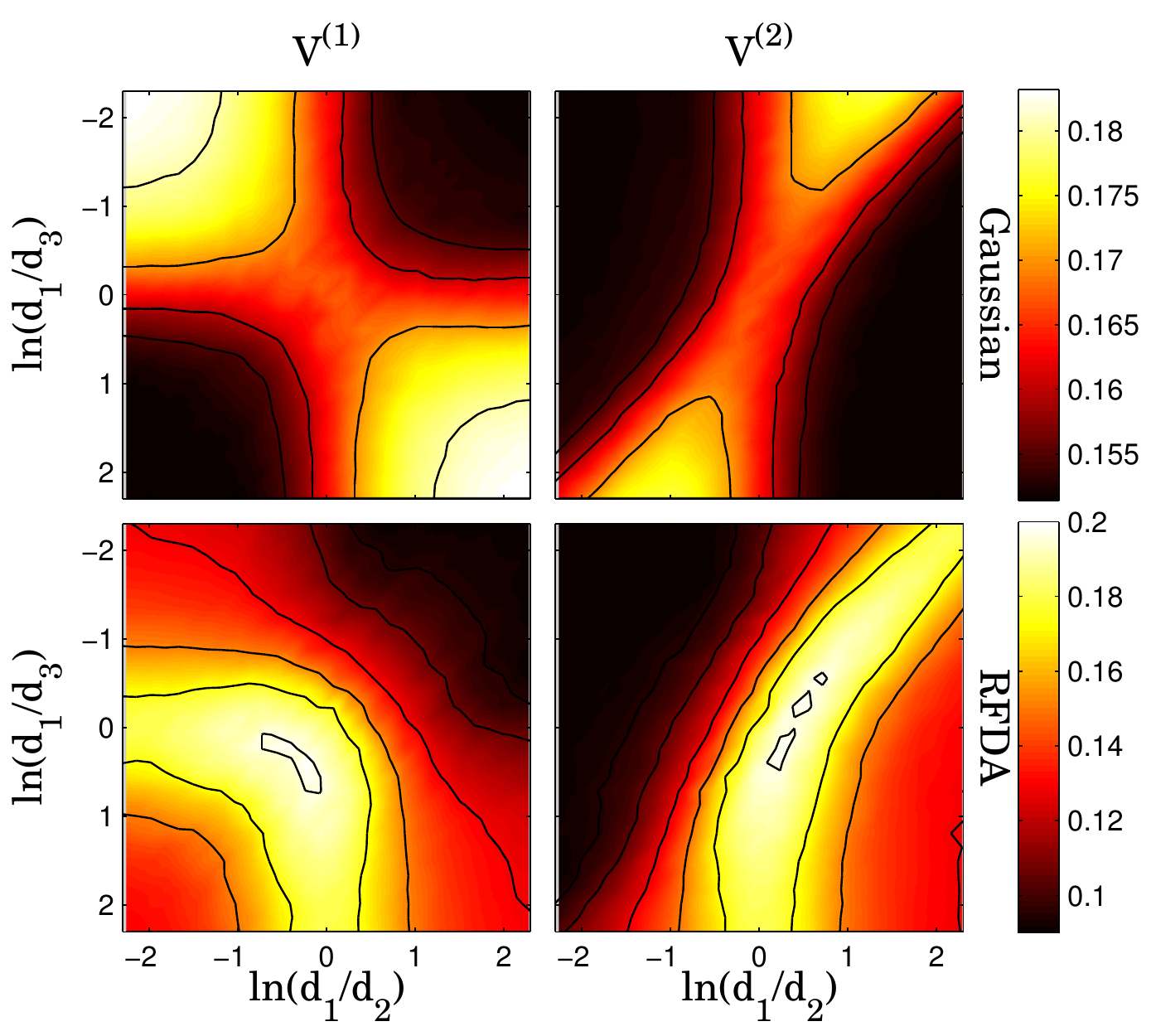,width=11cm}
\caption{\label{fig-variance-p-mods}  Variance of rate-of-change of the two ellipsoid orientation vectors ${\bf p}^{(1)}$ and  ${\bf p}^{(2)}$ as function of the ratios of semi-axes length. Top graphs: Gaussian stochastic model. Bottom graphs: results from RFDA Lagrangian stochastic model (see \S \ref{sec-RFDAmodel}). For the Gaussian case, contour lines correspond to values 0.155, 0.16, 0.17, 0.18, and for the RFDA case 0.09, 0.1, 0.12, 0.14, 0.16, 0.18, 0.2.}
\end{figure}

Simulation of this tensorial process generates a time series of ${\bf A}(t)$ which is used in the numerical solution of equation 
\ref{eq:JunkIllner}. The resulting variances of $\dot{\bf p}^{(1)}$ and $\dot{\bf p}^{(2)}$ are displayed in 
Fig. \ref{fig-variance-p-mods}, top two plots (the bottom ones show predictions from a model discussed below
in \S \ref{sec-RFDAmodel}). For the Gaussian process, it is seen that $V^{(1)}$ and $V^{(2)}$ depend only weakly on particle anisotropy, with values ranging only between 0.15 and 0.18. Further tests varying the forcing strength have been performed (not shown), and results are briefly commented upon in \S \ref{conclusions}.

\begin{figure}
\centering
\epsfig{file=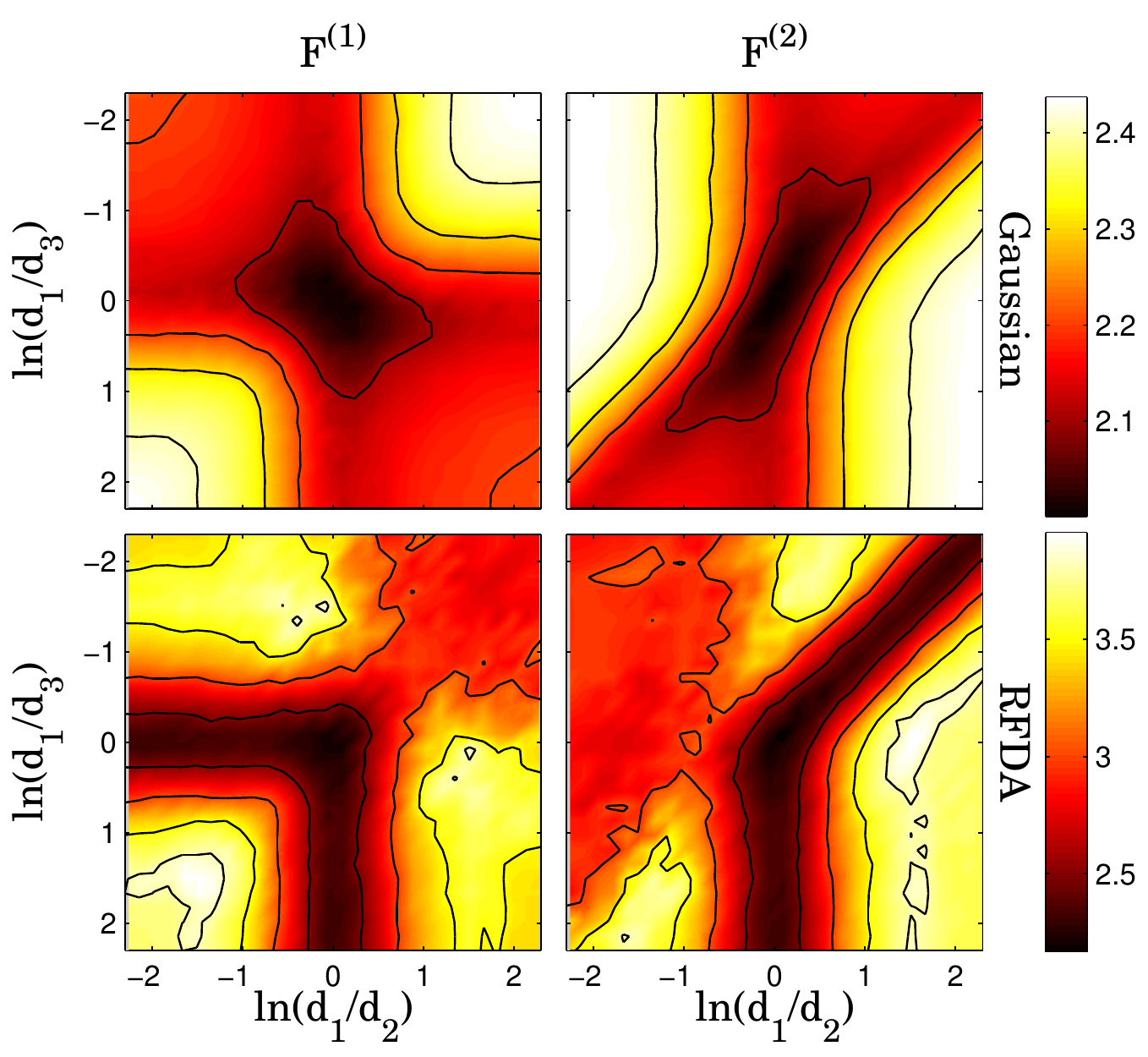,width=11cm}
\caption{\label{fig-flatness-p-mods}  Flatness of rate-of-change of the two ellipsoid orientation vectors ${\bf p}^{(1)}$ and  ${\bf p}^{(2)}$ as function of the two ratios of semi-axes length. Top graphs: Gaussian stochastic model, for which the flatness is near 2 for any ellipsoidal aspect ratios.  Bottom graphs: results from RFDA Lagrangian stochastic model (see \S \ref{sec-RFDAmodel}). For the Gaussian case, contour lines correspond to values 2.1, 2.2, 2.3, 2.4, and for the RFDA case, 2.2, 2.5, 3, 3.5, 3.8.}
\end{figure}

Results for the flatness $F^{(i)}$ (Eq. \ref{eq:Flatpi})  are shown in Fig. \ref{fig-flatness-p-mods} (top line). As can be seen, the Gaussian model (Eq. \ref{eq:Gaussian}) produces a flatness for the particle orientation rotation rates always close to 2 for all parameter values, at odds with DNS (c.f. Fig. \ref{fig-flatness-p-DNS}). The value of 2 can be exactly derived for spheres assuming orientation vectors independent from velocity gradients, themselves assumed Gaussian (see Appendix and Eq. \ref{eq:FlatSIspheres}). While the observed trends are difficult to distinguish from numerical noise, they also clearly differ from the 5/3 value obtained from assuming each element of ${\bf p}$ to be a Gaussian independent variable.   We are led to the conclusion that  assuming statistically independent $\textbf{A}$ and $\textbf{p}^{(i)}$, as well as a Gaussian process for $\textbf{A}$ with a correlation time-scale of $\tau_\eta$, gives a poor prediction of the rate of rotation of orientation vectors that has been observed in DNS.

Angular alignment trends will be discussed in the next section together with those of the RFDA model. 

\subsection{RFDA Lagrangian stochastic model}
\label{sec-RFDAmodel}

The  Lagrangian model developed in Ref. \cite{ChevillardMeneveau06} is given by the following non-dimensionalized, in units of the integral time scale $T$, stochastic differential equation
\begin{equation}\label{eq:RFDA}
d\textbf{A} =  \left(-\textbf{A}^2+
\frac{\mbox{Tr}(\textbf{A}^2)}{\mbox{Tr}(\textbf{C}_{{\tau_\eta/T}}^{-1})}
\textbf{C}_{{\tau_\eta/T}}^{-1}
-\frac{\mbox{Tr}(\textbf{C}_{{\tau_\eta/T}}^{-1})}{3}
\textbf{A}\right)dt+d\textbf{W}\mbox{ .}
\end{equation}
The \textit{recent Cauchy-Green tensor} $\textbf{C}_{\tau_\eta}$, which arises after invoking 
the \textit{recent fluid deformation} approximation (RFDA) \citep{ChevillardMeneveau06} is written in terms of  matrix exponentials as 
\begin{equation}\label{eq:MyCG}
\textbf{C}_{\tau_\eta} = e^{\tau_\eta \textbf{A}}e^{\tau_\eta
\textbf{A}^\top}\mbox{ ,}
\end{equation}
where $\tau_\eta$ is (as before) the Kolmogorov time-scale  (see \cite{Chevillardetal08} for additional details).
{The first three deterministic terms in Eq. \ref{eq:RFDA} represent, respectively, the exact self-stretching term, and models for both 
pressure Hessian and viscous diffusion. The modeling is based on a Lagrange-Eulerian change of variables 
coupled to the assumption that the Lagrangian pressure Hessian is an isotropic tensor. This differs from the assumption
of the restricted Euler model in which the Eulerian pressure Hessian is assumed to be isotropic. Assuming that the 
velocity gradient is perfectly correlated during a time $\tau_K$ along the Lagrangian trajectory, while it is uncorrelated for
longer time-delays, leads to a closed-form expression for the model pressure Hessian in the form of matrix exponentials.
The latter arise as solutions to the kinematic equation for the deformation tensor. A similar derivation can be
done for the Laplacian that arises in the viscous term.  An analysis of expansions of the matrix 
exponentials is provided in \cite{Martinsafonsomeneveau10}}.  The term $\textbf{W}$ is the same tensorial delta-correlated noise term that enters in the Gaussian process (Eq. \ref{eq:Gaussian}),  it represents possible forcing effects, e.g. from neighboring eddies. 

The RFDA model has been shown to reproduce several important characteristics of the velocity gradient tensor, such as the preferential alignments of vorticity with the intermediate eigen-direction of the strain {and subtle temporal correlations \citep{ChevillardMeneveau11}}. It has thus a more complex  covariance structure than the one obtained from the Ornstein-Uhlenbeck process (Eq. \ref{eq:Gaussian}) and is more realistic (see \cite{Chevillardetal08,Meneveau11}). Yet, the model has some known limitations: as discussed further in \cite{Meneveau11}, extensions to increasing Reynolds number (reducing $\tau_\eta/T$ below $10^{-2}$ or so) leads to unphysical tails in the velocity gradient PDFs. Also, tests have shown that the process leads to small deviations between the variance of the strain-rate tensor and the rotation rate tensor, i.e. for $\tau_\eta/T=0.1$, we obtain $\langle S_{ij} S_{ij}\rangle \approx 1.1 \langle \Omega_{ij} \Omega_{ij}\rangle$. Further strengths and limitations of the model will be highlighted by comparing its predictions of particle orientation dynamics to DNS. 

The process is simulated numerically using a standard second-order Runge-Kutta algorithm with a unique realization of the noise for each time step. The time series of ${\bf A}(t)$ generated by this process are, again,  used in the solution of equation 
\ref{eq:JunkIllner}. The resulting variances of $\dot{\bf p}^{(1)}$ and $\dot{\bf p}^{(2)}$ are displayed in 
the bottom row of plots in Fig. \ref{fig-variance-p-mods}. As can be seen, certain trends  agree well with
the results from the DNS. As opposed to the results from the Gaussian model, in the limit of rod-like particles
($d_1/d_2=d_1/d_3 \gg 1$), the variance of $\dot{\bf p}^{(1)}$ decreases significantly. Alignment of ${\bf p}^{(1)}$
with the vorticity leads to a reduction of its rate of change. In the other limit ($d_1/d_2=d_1/d_3 \ll 1$), however, the model  predicts also some reduction of variance, unlike the DNS results. To better understand the origin of this result, the alignments of ${\bf p}^{(1)}$ with the strain-rate eigensystem will be quantified and compared with DNS. 

In terms of the variance of $\dot{\bf p}^{(2)}$ shown in Fig. \ref{fig-variance-p-mods} (bottom right), we remark that there is good overall agreement between the model and DNS results: in the top left corner there is decreased variance, while towards the bottom-right corner the variance is generally higher. 
Nevertheless, a non-monotonic behavior is observed here too,  in which some decrease in variance towards $d_1/d_2\gg 1$ can be observed. 

\begin{figure}
\centering
\epsfig{file=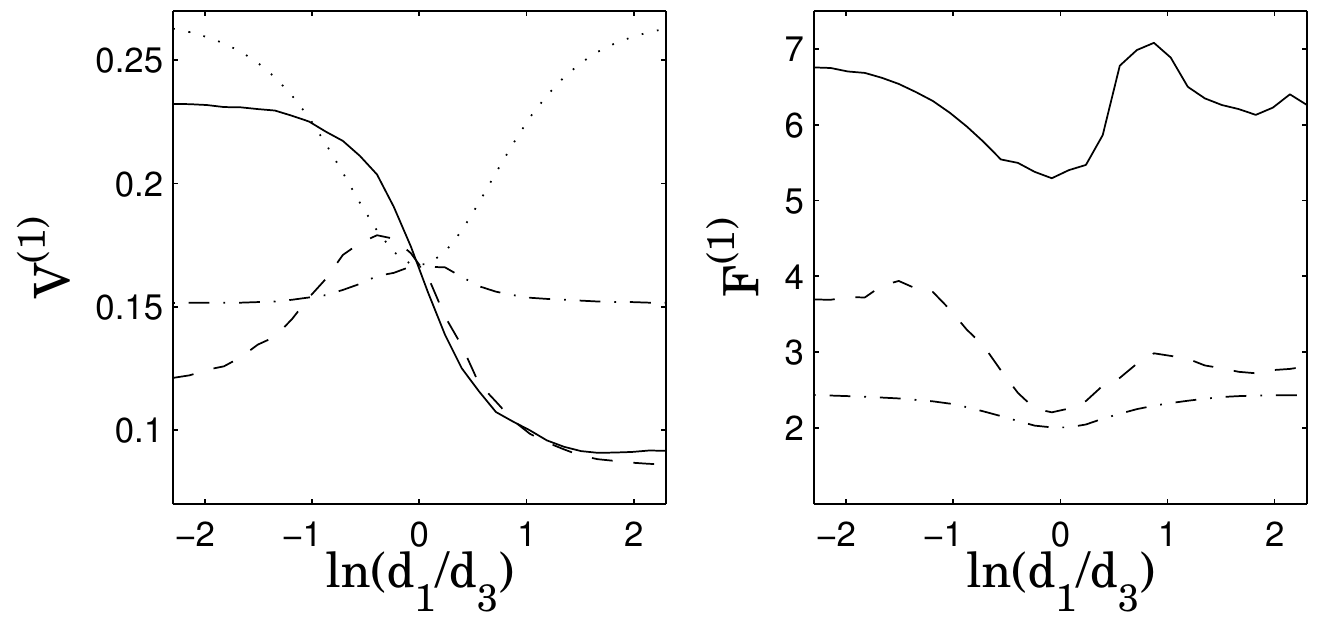,width=11cm}
\caption{\label{fig-variance-flatness-p-all}  Variance (left) and flatness factor (right) of  $\dot{\bf p}^{(1)}$ from DNS (solid line), the Gaussian model (dot-dashed), and the RFDA model (dashed line). We added also the prediction Eq. \ref{eq:ExactIndept} (dotted line) assuming that ${\bf p}^{(i)}$ and velocity gradients are statistically independent.}
\end{figure}

In order to enable quantitative comparisons between DNS, the Gaussian model (Eq. \ref{eq:Gaussian}), and the RFDA model (Eq. \ref{eq:RFDA}), we present sample results along the $d_1/d_2=d_1/d_3$ line of parameters, i.e. the axisymmetric cases.  Figure \ref{fig-variance-flatness-p-all} shows the variance of  $\dot{\bf p}^{(1)}$ for the DNS, Gaussian, and RFDA models. The first two lines are very similar to the results shown in \cite{Parsaetal12}. 

As far as the flatness factor is concerned (see Fig. \ref{fig-flatness-p-mods}, bottom row of contour plots), it is interesting to note that the  RFDA model predicts qualitatively quite well the rather complex trends observed in DNS and displayed in Fig. \ref{fig-flatness-p-DNS}. The main difference is that the flatness obtained from DNS is significantly higher that the one obtained from the RFDA model. This could be also related to the fact that the RFDA model under predicts the flatness of the velocity gradient elements themselves (see \cite{ChevillardMeneveau06}). We gather the results for the flatness of the rotation rate of orientation vectors obtained from DNS, Gaussian and RFDA models in Fig. \ref{fig-variance-flatness-p-all} (right panel). Indeed, we see that the RFDA follows accurately the variations observed in DNS, but the overall value (around 6.5 for DNS and around 3 for RFDA) is not reproduced.
\begin{figure}
\centering
\epsfig{file=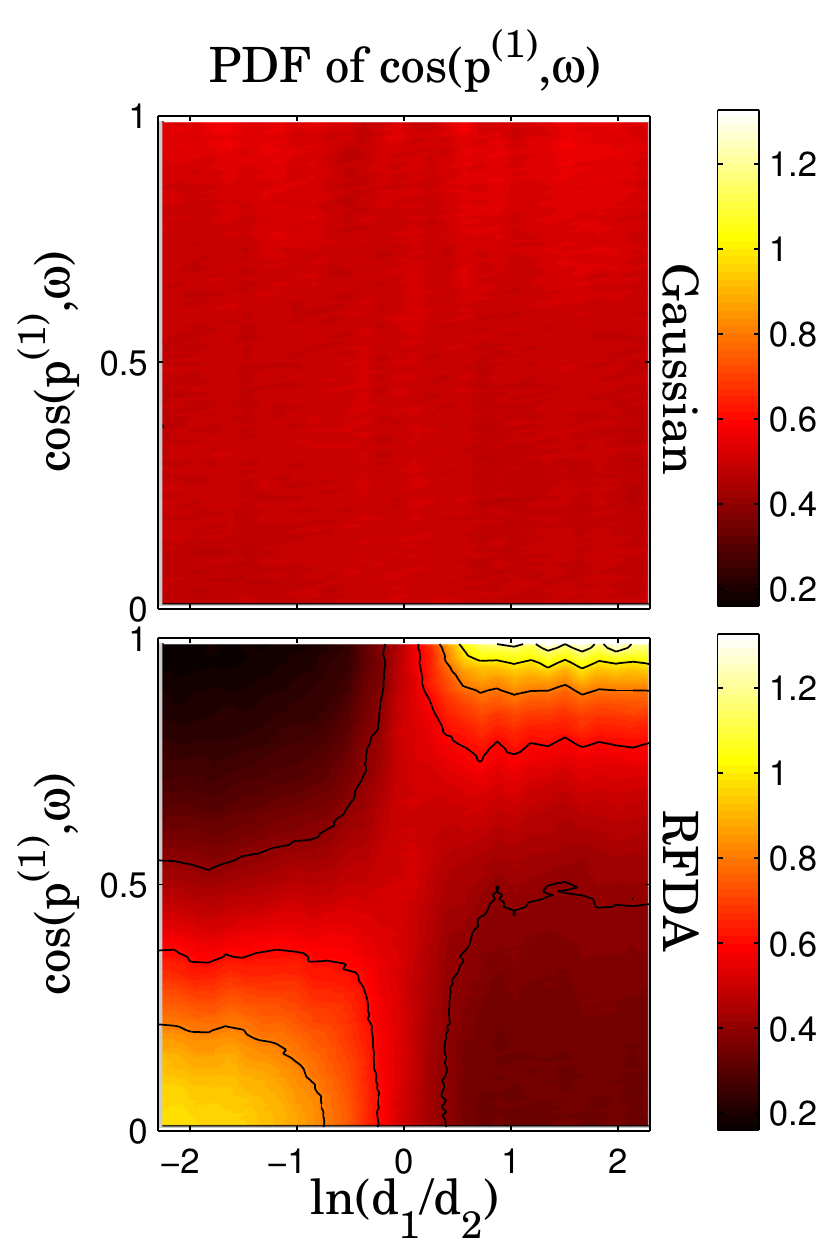,width=8cm}
\caption{\label{fig-align-pw-mods}  Probability-density of the cosine of the angle between the vorticity direction $\hat{\vct{\omega}}$ and the ellipsoid's major axis ${\bf p}^{(1)}$ for the axisymmetric case, as function of the anisotropy  parameter $\alpha = d_1/d_3=d_1/d_2$). Top row: Gaussian stochastic model (see \S \ref{sec-gaussian}), for which the alignment appears flat, for any ellipsoidal aspect ratios.  Bottom row: results from RFDA Lagrangian stochastic model. Contour lines correspond to values 0.4, 0.6, 0.8, 1, 1.2.}
\end{figure}

\begin{figure}
\centering
\epsfig{file=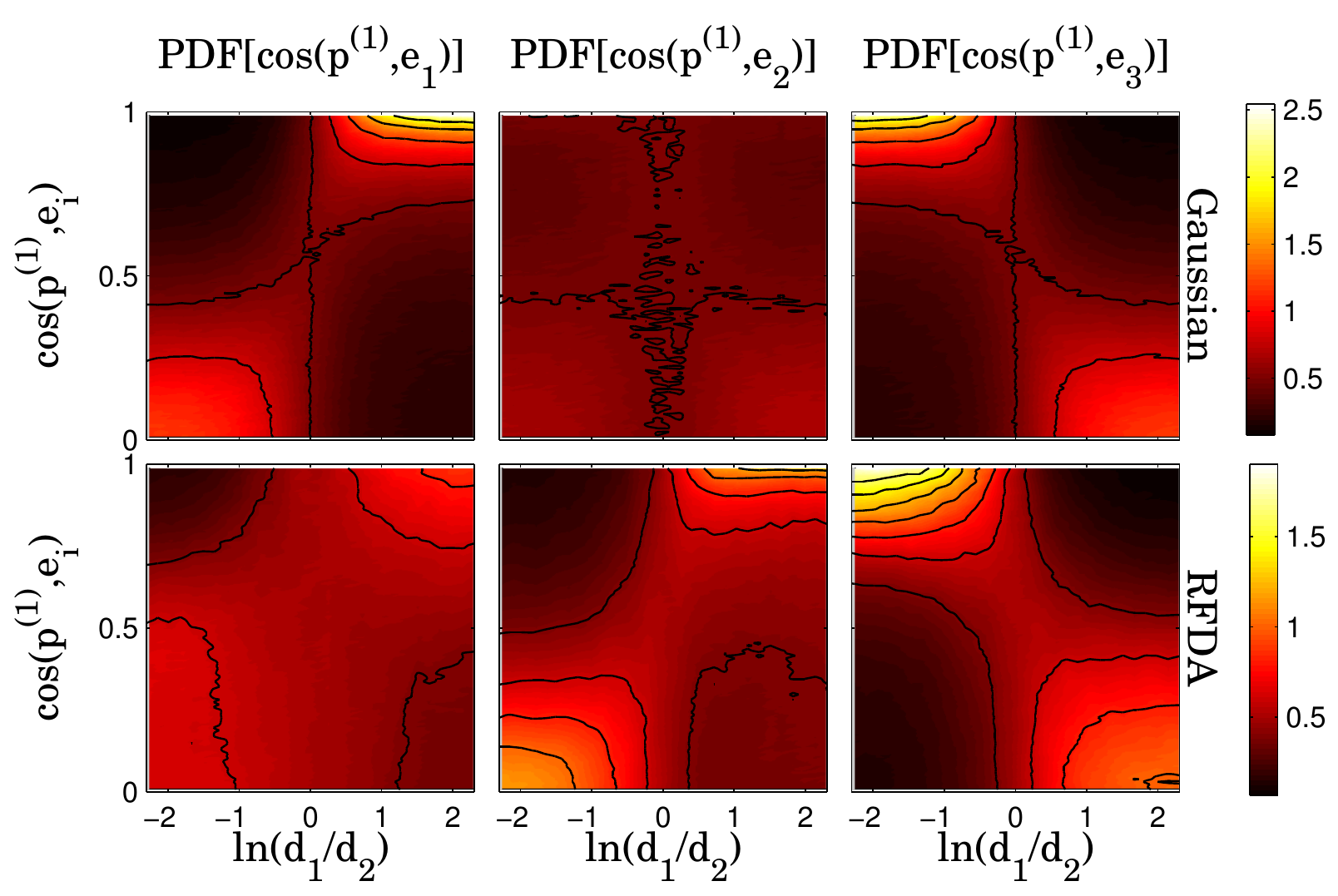,width=13.5cm}
\caption{\label{fig-align-pe-mods}  Probability-density of the cosine of the angle between the strain-rate tensor eigen-directions ${\bf e}_{i}$ and the ellipsoid's major axis ${\bf p}^{(1)}$ for the axisymmetric case ($\alpha = d_1/d_3=d_1/d_2$), as function of the anisotropy obtained from Gaussian (top line) and RFDA (bottom line) Lagrangian stochastic model. For the Gaussian case, contour lines correspond to values 0.5, 0.8, 1.2, 1.6, 2, and for the RFDA case, 0.4 0.6 0.8 1 1.2 1.4 1.6.}
\end{figure}

Analysis of model predictions of orientations of ${\bf p}^{(1)}$ for axisymmetric particles (i.e. $d_1/d_2=d_1/d_3$) with vorticity and strain-rate eigen-vectors
leads to the results shown in Figs. \ref{fig-align-pw-mods} and \ref{fig-align-pe-mods}, for both Gaussian and RFDA models.

For the Gaussian model, no preferential alignments of ${\bf p}^{(1)}$ with vorticity can be observed. Results for alignment with the   strain eigen-frame show a strong preferential alignment of fiber-like particles with the eigenvector associated to the most extensive eigen-direction, no preferential alignments with the intermediate eigen-direction, and preferential alignments of disc-like particles with the most contracting eigen-direction.  This numerical study reveals indeed, for this Gaussian model (Eq. \ref{eq:Gaussian}), a correlation between orientation vectors and velocity gradients, although of different nature as the one observed in DNS: whereas preferential alignments of fiber-like particles with intermediate eigen-direction and disc-like particle with most contracting one are found in DNS, Gaussian process only correctly predicts alignments of disc-like particles with most-contracting eigen-direction and reveals non realistic alignment properties of fibers. 

More refined Gaussian processes have been proposed for velocity gradient statistics. For instance,  \cite{PumirWilkinson11} and \cite{Vincenzi13} have considered  an Ornstein Uhlenbeck process  for $\textbf{A}$ with different correlation time scales for the symmetric and antisymmetric parts.  This is more realistic, since it is known that in turbulence the correlation time-scale for the rotation rate is significantly longer than that of the strain-rate. Applying this stochastic model to the orientation dynamics of rods (i.e. with $d_1/d_2=d_1/d_3 \to \infty$), \cite{PumirWilkinson11} observe similarly a strong preferential alignment of $\textbf{p}$ with the strain eigenvector associated to the most positive eigenvalue. This differs from the observations in DNS (see Figs. \ref{fig-align-pw-DNS} and \ref{fig-align-pe-DNS}), where  $\textbf{p}$ is instead found to be prefentially aligned with the direction of vorticity. As argued before, such trends then have immediate implications on the particle rotation rates. These results and arguments highlight the importance of both the temporal and alignment structure of $\textbf{A}$.  {Similar conclusions have been arrived at recently by \cite{GusEin13} who consider also the case of inertial axisymmetric particles and obtained analytical expressions for the rotation rate assuming an underlying Gaussian flow}. 

The alignments predicted by the RFDA model are significantly more realistic: 
As can be seen, for fibre-like particles ($d_1/d_2 \to \infty$), the model predicts strong alignment with vorticity (Fig. \ref{fig-align-pw-mods}, bottom row). In the other limit, for
disc-like particles, ${\bf p}^{(1)}$ in the model is preferentially perpendicular to the vorticity. That is to say, the vorticity is in the 
plane of the disc. Nevertheless, comparing in more detail with the DNS results in Fig. \ref{fig-align-pw-DNS}, it is evident that the model predicts 
a significantly broader (and weaker) alignment peak in the PDF (see different magnitudes given in colorbars). Hence, the alignment of the plane of the disc 
with vorticity is weaker in the model than in DNS, and as a result,  the variance of $\dot{\bf p}^{(1)}$ for discs (perpendicular to the disc plane) is reduced compared to the DNS result, where the alignment of the disc plane with vorticity is much sharper. 

\begin{figure}
\centering
\epsfig{file=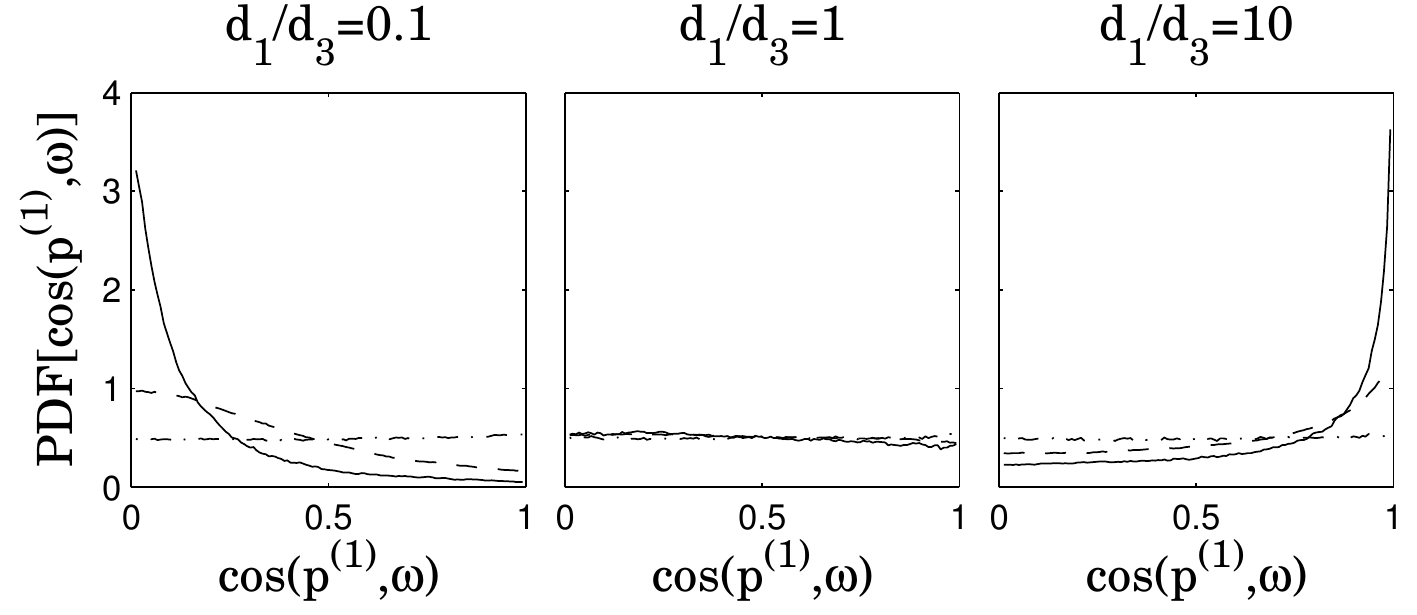,width=11.0cm}
\caption{\label{fig-alignments-pw-3a}  PDFs of the cosine of the angle between vorticity and ${\bf p}^{(1)}$, for three values of $\alpha$ along the 
$d_1/d_2=d_1/d_3$ line (the axisymmetric cases).  Results from DNS (solid line), Gaussian model (dot-dashed), and RFDA model (dashed line) are shown.}
\end{figure}

For better comparisons among DNS, the Gaussian model and the RFDA model, we present vorticity alignment result for three values of $d_1/d_3$ along the $d_1/d_2=d_1/d_3$  axisymmetry line.  Figure \ref{fig-alignments-pw-3a} shows the PDF of  
the cosine between ${\bf p}^{(1)}$ and the vorticity for the DNS, Gaussian, and RFDA model. We indeed see that, whereas the Gaussian model does not predict any preferential alignments, the RFDA model predicts correct trends. However, the preferential alignment (when $d_1/d_3 \to \infty$) and even more the orthogonality (when $d_1/d_3 \to 0$) are weaker than the ones observed in DNS.

The alignments of ${\bf p}^{(1)}$ with each of the strain-rate eigenvectors are also established using the PDF of the respective angle cosines. Results are shown in Fig. \ref{fig-align-pe-mods} (bottom line). 
Overall, the results from the RFDA model for the alignments of the orientation vectors with the strain-rate eigen-frame appears to be quite realistic, both in trends and amplitudes (compare Fig. \ref{fig-align-pe-DNS} and the bottom line of Fig. \ref{fig-align-pe-mods}). As in DNS, the strongest alignment trend seems to be that for disc-like particles to align with the most contracting eigen-direction.  As previously observed, rod-like particles tend to align well with the intermediate eigen-vectors, while disc-like particles show preponderance of perpendicular orientation with regards to the intermediate eigen-vector. Again, this is consistent with the trends with vorticity. Interestingly, alignment trends with the most stretching eigen-direction appear very weak, almost random, as in the DNS.

Finally, we have also considered the orientation  statistics of the vector ${\bf p}^{(2)}$ with respect to vorticity and strain-rate eigen-directions, along the $d_1/d_2 =(d_1/d_3)^{-1}$ line. The results (not shown) are very similar to those displayed in Figs. \ref{fig-align-pw-mods} and \ref{fig-align-pe-mods}, as it was observed for the DNS (Figs. \ref{fig-align-p2w-DNS} and \ref{fig-align-p2e-DNS}).

\section{Conclusions}
\label{conclusions}

In this study, the orientation dynamics of anisotropic particles in isotropic turbulence has been examined using DNS and some stochastic models. 
We have generalized a recent analysis for axisymmetric ellipsoids \citep{Parsaetal12} to the case of general, triaxial ellipsoidal tracer particles, without the assumption of
axisymmetry. The underlying evolution equation that has been used was developed using asymptotic analysis for inertia-free local linear flow by \cite{JunkIllner07}. The analysis is valid for small particles which are smaller than the turbulent flow's Kolmogorov scale. 

The orientation dynamics are characterized by the rates of rotation of the particle's  two perpendicular orientation unit vectors, their variances and flatness, as well as 
with distribution functions of the orientation with respect to the flow's vorticity and strain-rate eigen-vector directions. Measurements based on Lagrangian tracking in 
DNS of isotropic turbulence show that triaxial ellipsoids that are very long in one direction, very thin in another, and of intermediate size in the third direction exhibit reduced rotation rates that are similar to those of rods in the ellipsoid's longest direction. Conversely, they exhibit increased rotation rates that are similar to those of axisymmetric discs in the direction of their smallest thickness.

In comparing DNS results with various models and assumptions, we find that they differ significantly from the case when the particle orientations are assumed to be statistically independent from the velocity gradient tensor. DNS results also differ significantly from results obtained when the velocity gradient tensor is modeled using an Ornstein-Uhlenbeck  process with Gaussian statistics, with a forcing strength such that the velocity gradient standard deviation is on the same order of magnitude as the inverse of the temporal correlation time of the process. In that case, the velocity gradient tensor displays no  preferred vorticity-strain-rate tensor alignments. Further tests (not shown) with the Ornstein-Uhlenbeck process have been performed,  by varying the strength of the forcing. It is observed that when the forcing strength is reduced from the baseline (high) values (i.e. $\sim 1/\tau_\eta^{3/2}$) to values on the order of $1/T^{3/2}$ (or unity, when using units of $T$, as was the case for the RFDA model) 
while maintaining the correlation time fixed at $\tau_\eta$, then the results tend to those obtained assuming statistical independence among orientation vectors
and the velocity gradient tensor. The fact that the two approaches lead to the same result  
can be understood as follows: the dimensionless quantities $V^{(i)}$ may only depend on dimensionless parameters.
One of these is, e.g., $\theta = \tau_\eta \langle \Omega_{ij}\Omega_{ij} \rangle^{1/2}$,  a combination of the process correlation time-scale $\tau_\eta$ and the velocity gradient variance. Hence, reducing the forcing strength in the Gaussian process, i.e. letting $ \langle \Omega_{ij}\Omega_{ij} \rangle \to 0$  while keeping $\tau_\eta$ fixed implies $\theta \to 0$. This same limit may be achieved by keeping the variance fixed but reducing the correlation time-scale 
 $\tau_\eta \to 0$. When the correlation time-scale of the velocity gradient tensor tends to zero, one expects the same results as assuming statistical independence between the orientation vectors and the velocity  gradient tensor.  

DNS results are also compared with a stochastic model for the velocity gradient tensor in which the pressure and viscous effects are modeled based on the recent fluid deformation approximation (RFDA).  We remark that in the RFDA model, the nonlinear terms cause a large velocity gradient variance (finite $\theta$) even when the forcing strength is weak.  Thus in the RFDA model the variance and the correlation time are linked and cannot be independently controlled as they can be in the Ornstein-Uhlenbeck model.  Unlike the Gaussian linear model, the RFDA-based stochastic model accurately predicts the reduction in rotation rate in the longest direction of triaxial ellipsoids. This is due to the fact that this direction aligns well with the flow's vorticity, with its rotation perpendicular to the vorticity thus being reduced.  For disc-like particles, or in directions perpendicular to the longest direction in triaxial particles, the model predicts smaller rotation rates than those observed in DNS (although still larger than for rods). This behavior has been explained based on the probability of vorticity orientation with the most contracting strain-rate eigen-direction. In DNS, this alignment is very likely (sharp peak in the PDF), whereas the peak in the PDF predicted by the model is more diffused. {Furthermore,  the RFDA model falls short at reproducing the high flatness of the rotation rate amplitude (i.e. $p_ip_i$, see Fig. \ref{fig-variance-flatness-p-all} right panel), although trends (exceeding the Gaussian values) are consistent. The under-predition of flatness is likely due to the fact that the model does not reproduce the intermittent peak values of the velocity gradient components themselves (see \cite{ChevillardMeneveau06}) in which the high intensity tails of the velocity gradient elements can be seen to fall off faster than those in DNS.} Present results point to the need for further improvements in stochastic Lagrangian models for the velocity gradient tensor. Specifically, a model that predicts a sharper alignment between vorticity in a plane perpendicular to the most contracting strain-rate eigen-direction would be expected to lead to a more accurate prediction of the increased rotation rates of discs. 

{The deformation and breakup of viscous drops in shear flows  are greatly affected by the Lagrangian properties of fluid velocity gradients \citep{Stone94}.  
In some cases, it is possible to assume droplets are of ellipsoidal shape, as was done for example in \cite{MoslerShaqfeh97} and \cite{MaffettoneMinale98}, or even allowing for 
more non-trivial shape deformations \citep{Cristinietal03}. In either case, non-trivial correlations among the drop deformations and the strain eigendirections and vorticity of the flow suggest that turbulence will affect the rotation (tumbling) rates of deforming particles differently than is the case for rigid particles. Exploration of the RFDA model in the context of deforming particles is left for future studies.}

\vskip  1cm
Acknowledgements: The authors thank Emmanuel L\'ev\^eque for providing the authors with the Lagrangian time series of velocity gradient tensor from DNS, Bernard Castaing and Andrea Prosperetti for fruitful discussions, Denis Bartolo on plotting issues. CM thanks the Laboratoire de Physique de l'\'{E}cole Normale Sup\'{e}rieure de Lyon for their hospitality during a sabbatical stay and the US National Science Foundation (grant \# CBET-1033942) for support of turbulence research. We also thank the PSMN (ENS Lyon) for computational resources.

\appendix
\section{Calculation of the variance of rotation rates for the independent case}
\label{annex:Calc}
In this section, we present the calculation of $V^{(i)} = \langle \dot{\bf p}^{(i)}\cdot \dot{\bf p}^{(i)}\rangle$ for the case when it is assumed that ${\bf p}^{(i)}$, for any $i=1,2,3$, is statistically independent of 
the strain-rate and rotation tensors. We furthermore assume that each ${\bf p}^{(i)}$ is an isotropic vector, and recall that the set of vectors (${\bf p}^{(1)},{\bf p}^{(2)},{\bf p}^{(3)}$) is an orthonormal basis.  For this purpose, each side of the Junke-Illner equation (see Eq. \ref{eq:JunkIllnerIndex}) is squared and averaged:
$$
\langle \dot{p}_n^{(i)} \dot{p}_n^{(i)}\rangle = ~~~~~~~~~~~~~~~~~~~~~~~~
$$
\begin{equation}\label{eq:averagepdoti}
 \left<  \left( \Omega_{nj} p^{(i)}_j +\sum \limits_{k,m} \epsilon_{ikm} \lambda^{(m)} p^{(k)}_n p^{(k)}_q S_{ql} p^{(i)}_l \right) 
                                                                                              \left( \Omega_{ng}p^{(i)}_g+\sum \limits_{a,b} \epsilon_{iab} \lambda^{(b)}   p^{(a)}_n p^{(a)}_r S_{rs} p^{(i)}_s \right) \right>.
\end{equation}
Expanding and using the assumption of statistical independence between ${\bf p}$, 
${\bf S}$ and $\vct{\Omega}$, the following expressions must be evaluated:
\begin{equation}\label{eq:Firstterm}
\langle \Omega_{nj} p^{(i)}_j \Omega_{ng}p^{(i)}_g \rangle = \langle \Omega_{nj}  \Omega_{ng}\rangle ~  \langle p^{(i)}_j p^{(i)}_g \rangle,
\end{equation}
\begin{equation}\label{eq:ContribLin}
2\langle \Omega_{nj} p^{(i)}_j \sum \limits_{a,b} \epsilon_{iab} \lambda^{(b)}   p^{(a)}_n p^{(a)}_r S_{rs} p^{(i)}_s \rangle = 
2\langle \Omega_{nj}  S_{rs} \rangle \sum \limits_{a,b} \epsilon_{iab} \lambda^{(b)} \langle  p^{(i)}_j p^{(i)}_s  p^{(a)}_n p^{(a)}_r \rangle,
\end{equation}
and
$$
\langle  \sum \limits_{k,m} \epsilon_{ikm} \lambda^{(m)} p^{(k)}_n p^{(k)}_q S_{ql} p^{(i)}_l  \sum \limits_{a,b} \epsilon_{iab} \lambda^{(b)}   p^{(a)}_n p^{(a)}_r S_{rs} p^{(i)}_s   \rangle = 
$$
\begin{equation}\label{eq:6term}
\langle S_{ql}   S_{rs} \rangle ~  \sum \limits_{k,m} \sum \limits_{a,b}  \epsilon_{ikm} \epsilon_{iab} \lambda^{(m)}  \lambda^{(b)} 
 \langle   p^{(k)}_n  p^{(a)}_n p^{(k)}_q  p^{(a)}_r p^{(i)}_l  p^{(i)}_s  \rangle.
\end{equation}
The set of vectors (${\bf p}^{(1)},{\bf p}^{(2)},{\bf p}^{(3)}$) is an orthonormal basis, thus
$$ p^{(k)}_np^{(a)}_n = \delta_{ka}.$$
This implies that Eq. \ref{eq:6term} simplifies to
$$
\langle  \sum \limits_{k,m} \epsilon_{ikm} \lambda^{(m)} p^{(k)}_n p^{(k)}_q S_{ql} p^{(i)}_l  \sum \limits_{a,b} \epsilon_{iab} \lambda^{(b)}   p^{(a)}_n p^{(a)}_r S_{rs} p^{(i)}_s   \rangle = 
$$
\begin{equation}\label{eq:4term}
\langle S_{ql}   S_{rs} \rangle ~  \sum \limits_{m,b,k} \epsilon_{ikm} \epsilon_{ikb} \lambda^{(m)}  \lambda^{(b)} 
 \langle  p^{(k)}_q  p^{(k)}_r p^{(i)}_l  p^{(i)}_s  \rangle.
\end{equation}
We can see that the average of $ |\dot{{\bf p}}^{(i)}|^2$ (Eq. \ref{eq:averagepdoti}) depends only on the statistics of velocity gradients and second and fourth moments of orientation vector components. 
Assumption of isotropy for the unit-vectors $\dot{{\bf p}}^{(i)}$ implies that:
\begin{equation}\label{eq:averagenormiso}
 \langle p^{(i)}_j p^{(i)}_g \rangle = \frac{1}{3} \delta_{jg}, 
\end{equation}
\begin{equation}\label{eq:IsoForm4}
\langle  p^{(k)}_q  p^{(k)}_r p^{(i)}_l  p^{(i)}_s  \rangle  = A^{(k,i)} \delta_{qr}\delta_{ls}
  + B^{(k,i)}\delta_{ql}\delta_{rs} + C^{(k,i)}\delta_{qs}\delta_{rl}.
\end{equation}
From there it is easily seen that Eq. \ref{eq:ContribLin} gives no contributions since any contractions of $\Omega_{nj}  S_{rs}$ vanish. The 3 remaining unknown coefficients that enter in the evaluation of Eq.  \ref{eq:4term} may be found by specifying particular values. For instance, the index contraction $q=r$ and $l=s$ yields
\begin{equation}
  \langle  p^{(k)}_q p^{(k)}_q  p^{(i)}_l p^{(i)}_l \rangle =  \langle  1 \times 1 \rangle = 1 =   9 A^{(k,i)}
  + 3 B^{(k,i)} +3  C^{(k,i)}.
\end{equation}
Inspecting Eq. \ref{eq:4term}, we notice that terms in the sum such that $k=i$ give no contribution because of $\epsilon_{ikm}\epsilon_{ikb}$. Thus, we consider $k\ne i$. In this case, the contraction $q=l$ and $r=s$ yields:
\begin{equation}
0= 
 3 A^{(k,i)}  + 9 B^{(k,i)}+ 3 C^{(k,i)},
 \end{equation}
and the contraction $q=s$ and $r=l$ yields:
 \begin{equation}
0= 3 A^{(k,i)}   + 3  B^{(k,i)}  + 9 C^{(k,i)}.
\end{equation}
Solving these equations yields in the case $k\ne i$
\begin{equation}
A^{(k,i)} = \frac{2}{15} ~~~~{\rm and} ~~ B^{(i,a)}=C^{(i,a)} = -\frac{1}{30}.
\end{equation}
Finally, simplifying Eq. \ref{eq:Firstterm} with Eq.  \ref{eq:averagenormiso} and contracting the isotropic form Eq. \ref{eq:IsoForm4} with $\langle S_{ql}   S_{rs} \rangle $, we get
$$ \langle |\dot{{\bf p}}^{(i)}|^2 \rangle = \frac{1}{3} \langle \Omega_{pq}\Omega_{pq}\rangle +\frac{1}{10}\langle S_{pq}S_{pq}\rangle \sum \limits_{m,b,k\ne i} \epsilon_{ikm} \epsilon_{ikb} \lambda^{(m)}  \lambda^{(b)}.$$
Using the isotropic relations $ \langle \Omega_{pq}\Omega_{pq}\rangle  = \langle S_{pq}S_{pq}\rangle$ and normalizing the former relation by $ 2\langle \Omega_{pq}\Omega_{pq}\rangle = \varepsilon/\nu$, we get the following functional forms for the fluctuation of rotation rate variances:
\be\label{eq:finalresAnn}
V^{(i)}_{\rm SI} = \frac{1}{6} +\frac{1}{20} \sum_{k\ne i}\left(\lambda^{(k)}\right)^2.
\ee

If furthermore the statistics of $\textbf{A}$ are assumed Gaussian (as in Section \ref{sec-gaussian}), it is also possible to derive exactly the value for the flatness, although the calculation is more tedious. For the particular case of spheres, i.e. $d_1=d_2=d_3$ or $\lambda^{(i)}=0$, the dynamics is rather simple since $\dot{p}_n^{(i)}=\Omega_{nj} p^{(i)}_j$. Assuming $\textbf{p}^{(i)}$ isotropic and independent on $\textbf{A}$, we easily get $\langle |\dot{\textbf{p}}^{(i)}|^4\rangle =\frac{1}{15}[\langle \mbox{tr}^2 \vct{\Omega}\vct{\Omega}^\top\rangle+2\langle \mbox{tr} (\vct{\Omega}\vct{\Omega}^\top)^2\rangle]$. For isotropic, homogeneous and trace-free Gaussian velocity gradient tensors, we get $\langle \mbox{tr}^2 \vct{\Omega}\vct{\Omega}^\top\rangle =\frac{5}{3}\langle \mbox{tr} \vct{\Omega}\vct{\Omega}^\top\rangle^2$ and $\langle \mbox{tr} (\vct{\Omega}\vct{\Omega}^\top)^2\rangle =\frac{5}{6}\langle \mbox{tr} \vct{\Omega}\vct{\Omega}^\top\rangle^2$. Since, $\langle |\dot{\textbf{p}}^{(i)}|^2\rangle =\frac{1}{3}\langle \mbox{tr} \vct{\Omega}\vct{\Omega}^\top\rangle$, we finally get:
\begin{equation}\label{eq:FlatSIspheres}
F^{(i)}_{\rm SI}(1,1) = 2.
\end{equation}

 

\end{document}